\documentclass[aps,pra,preprint,superscriptaddress, showpacs]{revtex4}
\usepackage{graphicx}
\usepackage{subfigure}
\usepackage{amsmath}
\begin{document}


\title{Modelling of optical traps for aerosols}


\author{D. R. Burnham}
\altaffiliation{Current address: Department of Chemistry, University of Washington, Box 351700, Seattle, WA 98195-1700}
\affiliation{SUPA, School of Physics and Astronomy, University of St Andrews, North Haugh, Fife, KY16 9SS, UK}
\affiliation{SUPA, Electronic Engineering and Physics Division, University of Dundee, Nethergate, Dundee, DD1 4HN, UK}
\author{D. McGloin}
\email[]{d.mcgloin@dundee.ac.uk}
\affiliation{SUPA, Electronic Engineering and Physics Division, University of Dundee, Nethergate, Dundee, DD1 4HN, UK}


\date{\today}

\begin{abstract}
Experimental observations suggest that there are differences between the behavior of particles optically trapped in air and trapped in a liquid phase. We present a modified version of Mie Debye Spherical Aberration theory to numerically simulate such optical system in attempt to explain and predict these effects. The model incorporates Mie scattering and focussing of the trapping beam through media of stratified refractive index. Our results show a geometrical optics approach cannot correctly describe our system and that spherical aberration must be included. We successfully qualitatively explain the observed phenomena and those of other authors, before discussing the limits of our experimental techniques and methods to improve it. We draw the important conclusion that when optically trapping aerosols the system does not behave as a true `optical tweezers', varying between levitation and single beam gradient force trapping depending on particle and beam parameters.
\end{abstract}


\maketitle

\section{Introduction}
Optical manipulation has matured considerably in recent years to become a powerful quantitative tool~\cite{Pesce2005,Ghislain1994}. The forces imparted are ideally situated to enable them to act as force transducers in molecular biology~\cite{Lang2002}. Also their ability to trap and isolate single or multiple objects makes them powerful in many different disciplines including biochemistry~\cite{Kuyper2003}, fluidics~\cite{Di2006}, condensed matter and fundamental physics~\cite{Hertlein2008,McCann1999}. One new application area, which this paper discusses, is the use of optical traps for studying aerosols and their basic physical and chemical properties~\cite{Mitchem2008}.

In all these disciplines it is useful to have an understanding of the processes occurring including the optically generated forces, which clearly underpin many of the measurments being undertaken. Modelling of optical forces is already used extensively in the field of optical manipulation to understand, for example, force mapping~\citep{Zhao2008} and optical binding~\citep{Metzger2006a} or to extract physical parameters not otherwise obtainable from measurements~\citep{Knoner2006}. One of the best examples is the understanding of how optical forces affect the cytoskeleton. Changes in the elasticity of this matrix are directly related to stages of cancer within individual cells~\citep{Guck2005}. As we will see, when optically trapping aerosols, we are working at the very edge of what can be considered to be `optical tweezers' and so the modelling of the forces involved may act as a method for testing the theories at their limits.

Aerosols are a significant constituent of the atmosphere and a major factor in determining its chemical balance, for example the ozone hole and acid rain~\cite{Reid2005}, impairing visibility and contributing to radiative balance~\cite{Jacobson2002}. Furthermore, understanding aerosol behaviour, how they enter and interact with the body~\cite{Jacobson2002} is relevant for both the effects of pollution on the human population and the efficacy of medicinal drugs~\cite{Labiris2003}.

Aerosols can be grouped into three main categories or modes. The nucleation mode consists of small emitted or newly nucleated particles with a mean radius less than $0.05~\mu\text{m}$. Upon growth and coagulation these particles move into the accumulation mode with radii between $0.05~\mu\text{m}$ and $1~\mu\text{m}$. Finally the coarse mode of aerosols consists of particles with radii greater than $1~\mu\text{m}$~\cite{Jacobson2002}.

In terms of surface area to volume ratio, accumulation mode aerosol constitutes the largest proportion of atmospheric aerosol and therefore dominates atmospheric aerosol chemistry~\cite{Reid2005}. This mode is also able to penetrate deep into the lungs playing a major role in the interaction of aerosol with the human body. Its near optical wavelength size also affects visibility~\cite{Jacobson2002}. Most optical manipulation work to date concentrates on coarse mode liquid phase aerosols, particularly relevant in drug delivery and atmospheric chemistry. However, the questions that can be answered with current techniques are limited so they will need to be altered to allow accumulation mode aerosol to be studied.

Our study aims to investigate the optical forces imparted to optically trapped airborne microscopic spheres. This is a very difficult task to carry out experimentally~\citep{Smith2003a} and the standard technique is to computationally model momentum transfer from focussed beam to particle.

Our aim in developing models examining the optical forces involved in airborne optical traps is to increase the understanding of the outcomes from relevant experiments to give a more complete picture of the process. We can also probe the boundaries of the current experiment to see if they can be extended.

Symmetry within the system simplifies the mathematics. A sphere in an axially symmetric beam is probably the simplest of formulations, with a large amount of the constituent work already available. Should one like to model the trapping of non-spherical objects then the computation becomes more complex with the T-matrix approach becoming the favoured method. Once the T-matrix has been calculated for a given object it need not be calculated again for every orientation of the object in the beam, thus making it rather advantageous. Nieminen \textit{et al.}~\citep{Nieminen2001a} have used this approach to code an `optical tweezers computational toolbox' freely available for use~\citep{Nieminen2007}.

The principle problems with many of the approaches available are the over-complexity (ours is a relatively simple problem), their inappropriateness for the size scale we are looking at, and the lack of a description of the true trapping beam profile, as we will discuss shortly. A lucid introduction to the inadequacies is given by Viana \textit{et al.}~\citep{Viana2007}. The microdroplets we are studying are $\gtrsim1~\mu\text{m}$ in radius so the force calculation lies above the Rayleigh regime~\cite{Ashkin1986} and to a first approximation can be described by a geometrical optics (GO) model~\cite{Roosen1977}. It will quickly become apparent that by studying airborne objects a GO model cannot deal with the system under study. To resolve this problem we implement a model that uses an integral representation of focussed light crossing refractive index interfaces and an exact form of plane wave scattering from spheres.

We have previously shown~\cite{BurnhamBrownian} that the motion of optically trapped droplets can be described using a simple harmonic oscillator model. Evidence also suggests that the odd phenomena observed in the optical trapping of aerosols originates from the water droplets' interaction with the optical field. The principle aim of this investigation is to see if the isolated physics of optical forces leads to the phenomena observed in experiments and to see how far the boundaries of the techniques can be pushed. We will begin by stating the experimental phenomena observed by ourselves and others:
\begin{enumerate}
\item As trapping laser power increases so does the height above the water layer or coverslip that the droplet is trapped~\cite{Knox2007}.
\item With further increases of power the droplet is lost from the trap. This does not always occur and is more pronounced for smaller droplets~\cite{BurnhamBrownian}.
\item After first capture, the droplet can undergo significant growth or evaporation coupled with large axial oscillations. These oscillations can occur significantly after capture but are far slower~\cite{McGloin2008,Di2007a}.
\item There is a linear dependence of `captured' droplet radius with trapping power. Small droplets cannot be trapped with high power~\cite{Hopkins2004,Burnham2006}.
\end{enumerate}
The boundaries we would like to determine and push are;
\begin{itemize}
\item the range of particle refractive indices it is possible to trap,
\item the limits of particle radius that can be trapped with the current apparatus.
\end{itemize}

Fortunately our system is highly symmetric and almost complete rigorous wave theory solutions already exist in literature, although no computer code is readily available and modifications must be made to suit our problem.

\begin{figure}[!ht]
	\begin{center}
	\subfigure[]{\includegraphics[width=8.6cm]{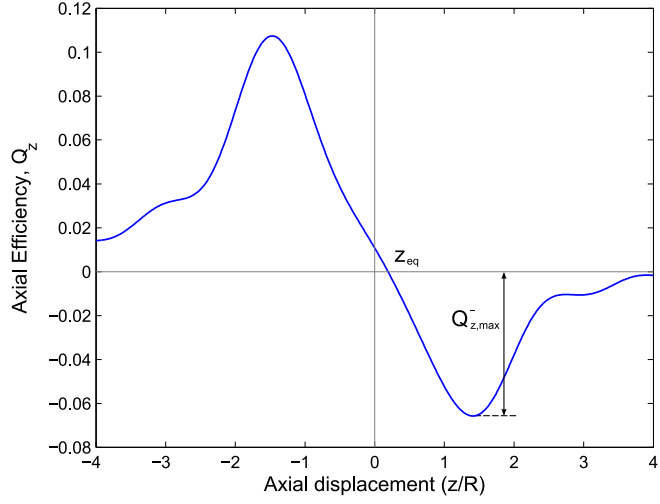}\label{Fig:Axial}}
	\\
	\subfigure[]{\includegraphics[width=8.6cm]{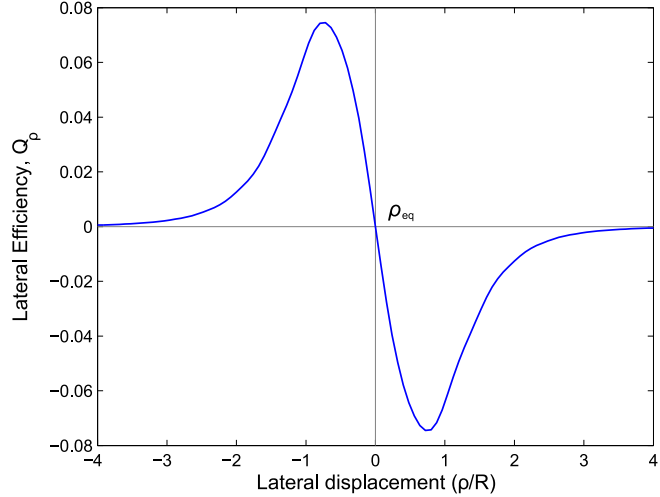}\label{Fig:Lateral}}
	\caption{(Color online) (a) Axial efficiency, $Q_{z}$, as a function of axial displacement. The axial equilibrium position, $z_{eq}$, occurs where the curve crosses y = 0 with negative gradient, $Q_{z,max}^{-}$ quantifies the traps axial strength. (b) Lateral efficiency, $Q_{\rho}$, as a function of lateral position. The lateral equilibrium position, $\rho_{eq}$, occurs where the curve crosses y = 0 with negative gradient. For (a) and (b) the gradient of the curve at $z_{eq}$ and $\rho_{eq}$ are proportional to the axial and lateral trap stiffness, $\kappa_{z}$ and $\kappa_{\rho}$ respectively.}
	\label{Fig:Output}
	\end{center}
\end{figure}

Here the forces exerted on spheres are decomposed into two directions, lateral and axial. That is the direction perpendicular to and crossing the beam propagation axis, and the direction lying on the beam propagation axis respectively. The simulations programmed in MATLAB calculate, for a given point on one of the two axes described, the efficiency with which momentum is transferred to the object, $Q$. The optical force, $F$, can be determined through
%
\begin{equation}
	F=Q\frac{n_{m}P}{c},
	\label{eq:momentumefficiency}
\end{equation}
where $P$ is the trapping power, $n_{m}$ is the refractive index of the medium, and $c$ is the speed of light. We are interested in observing how force varies with position along the axis and as such output force curves that are either a function of lateral, or axial displacement. From these force curves several parameters can be taken or calculated that describe the system under study as re-illustrated in fig.~\ref{Fig:Output}~\citep{Stilgoe2008}.

An object placed at a point where it experiences zero total force and is surrounded (within a certain proximity) by a negative gradient is said to be in \textit{equilibrium}. Should the object be displaced the local gradient will produce a restoring force back toward this equilibrium position. In figures~\ref{Fig:Axial} and~\ref{Fig:Lateral} these positions, $z_{eq}$ and $\rho_{eq}$, are the axial and lateral equilibrium positions respectively.

To achieve a single beam gradient force trap the efficiency, Q, in the negative axial direction must overcome that opposing it. If accomplished the efficiency curve will at some position become negative, allowing a point of zero force and negative gradient to exist. The maximum magnitude of this, $Q_{z,max}^{-}$, is a good measure of the axial optical trap strength~\citep{Stilgoe2008}, with its sign indicating whether a stable equilibrium position exists or not.

The axial and lateral trap stiffness is proportional to the gradient of the force curves at their equilibrium position, which for the lateral case, assuming symmetry, is at zero. The trap stiffness, $\kappa$, is related to the efficiency, $Q$, by
\begin{equation}
	\kappa = -\frac{n_{m}P}{c}\frac{\partial Q}{\partial s}.
	\label{eq:stiffness}
\end{equation}
where $s$ is either $z$ or $\rho$ for axial and lateral respectively.

During the results and discussion we shall look at the force curves alone and also results that are determined from many such curves where the parameters have been varied.

\section{The problem}
\begin{figure}[!ht]
	\begin{center}
	\subfigure[]{\includegraphics[width=4cm]{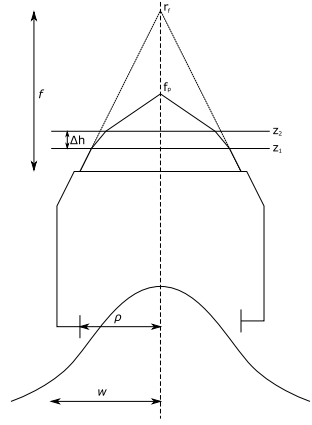}}
	\\
	\subfigure[]{\includegraphics[width=4cm]{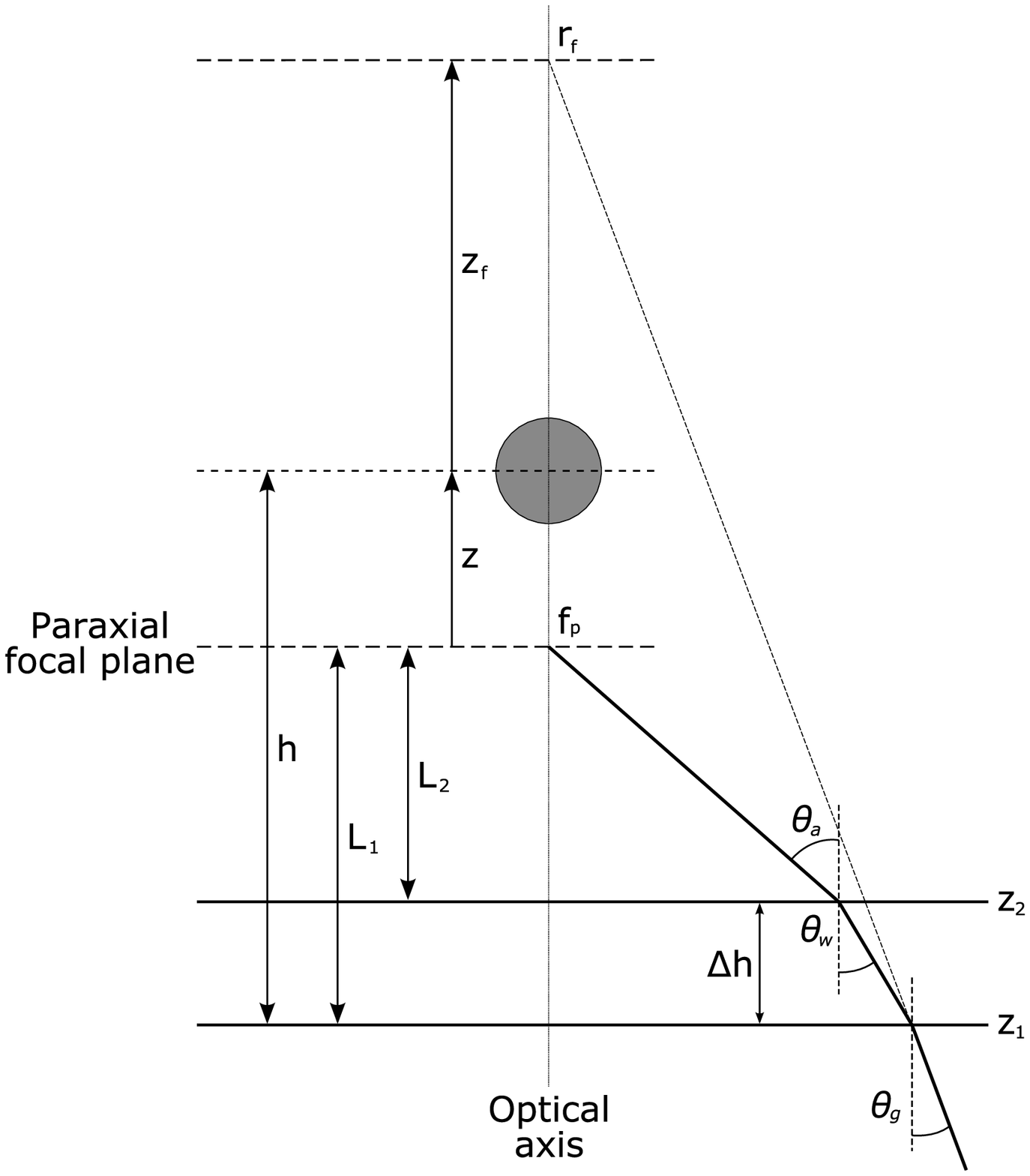}}
	\caption{Optical system and its parameters. a) A beam of wasit $w$ enters an objective lens of focal length $f$ with a back aperture of radius $\rho$. It is focussed to a point $f_{p}$ having propagated through two mismatched refractive index interfaces, $z_{1}$ and $z_{2}$, such that the thickness of the middle medium is $\Delta h$. If neither interface existed then the light would be focussed to point $r_{f}$. b) Expanded view of the focal region of the microscope objective to the left. Light is incident on the first interface, $z_{1}$, at an angle $\theta_g$ and refracted to an angle $\theta_{w}$. It is then incident on the second interface $z_{2}$, a distance $\Delta h$ away, at an angle $\theta_{w}$ where it is refracted to an angle $\theta_{a}$ and focussed to its paraxial focus point $f_{p}$. The height of the paraxial focus above the second and first interfaces is $L_{2}$ and $L_{1}$ respectively. The droplet is trapped a distance $h$ above the first interface, $z$ above the paraxial focus and $z_{f}$ below the point $r_{f}$. The first interface is between glass and water, and the second interface is between water and air.}
	\label{fig:Wolf}
	\end{center}
\end{figure}

Ideally we want to model forces on a microdroplet trapped near the focus of a beam given only the properties readily known. The geometry of our system is significantly different to those normally studied~\cite{Joykutty2005,Keen2007} as an additional discrete step in refractive index is located between the microscope objective and trapped particle due to the method of trap loading~\cite{Burnham2006,Mitchem2008}. The detailed system is illustrated in fig.~\ref{fig:Wolf} with a laser beam of wavelength $\lambda$, waist $w$, incident upon the back aperture of an objective lens of focal length $f$, aperture $\rho$ and focussed to the diffraction limit at a converging angle $\theta_{0}$, normally quantified in terms of numerical aperture, $\text{NA}=n_{m}\sin{\theta_{0}}$.

Our system assumes matching of objective lens, oil and glass coverslip refractive indices resulting in three consecutive layers of media and two refractive index mis-matched interfaces.


For the problem stated the decision of which size regime in which to calculate optical forces is important. For the particle radii here, $R\simeq\lambda$, we are outside the Rayleigh regime and to a first approximation a GO approach would seem appropriate. However, considering the problem we see that this is not the case for the following reasons.

Firstly, the NA, hence opening angle $\theta_{0}$, of optical traps is large and the paraxial assumption ($\sin{\theta}\simeq\theta$) is no longer applicable. For highly convergent beams the focus is not Gaussian but rather governed by an integral representation due to the electromagnetic diffraction within the optical system~\citep{Richards1959,Wolf1959}. Also, as fig.~\ref{fig:Wolf} shows the beam is focussed through a coverslip and an aqueous layer. The interfaces created by this glass slide to aqueous layer and aqueous layer to trapping medium (air) creates a mismatch in refractive index through which the beam is focussed. These discontinuities introduce spherical aberration into the focussed beam and so can only be fully described using a full wave analysis.

Secondly, the interaction of a plane wave with a sphere where $R\simeq\lambda$ is more challenging to describe than by constructing the problem as a beam of many single rays passing through a sphere, as in GO. The description really must take into account diffraction. There is an analytical solution to Maxwell's equations for the scattering of a plane light wave by a single sphere for any ratio of radius to wavelength. The solution was independently developed by Mie, Debye, and Lorenz around the turn of the 20th century but has been historically referred to as `Mie theory' or `Mie scattering'~\citep{Kerker1969}.

Thirdly, there is something intuitively wrong with the wavelength independence of GO. The focal waist and scattering of light by colloidal particles is known experimentally to be wavelength dependent so a theory of optical tweezers should also be wavelength dependent. An additional complexity arises from the proximity of reflecting surfaces which can cause `reverberations'~\citep{Viana2007} of light that significantly alter the interaction.

Finally, a true description of the physics at play must traverse the full range of applicability from Rayleigh scattering to GO.

\section{Theory}
We will initially utilise results from a GO approach to describe the forces on spheres to enable a comparison to a large number of relevant articles already in literature~\cite{Gong2007,Gussgard1992,Kohira2005}. We chose to use the GO model, described in detail by Mazolli \textit{et al.}, which takes into account the Gaussian nature of the beam and the Abbe sine condition~\cite{Mazolli2003}.

It will become apparent that theories beyond GO are needed to accurately model our problem. For this purpose we rely heavily on the Mie-Debye model given by Mazolli \textit{et al.}~\cite{Mazolli2003} and its extension to Mie-Debye-Sphercal-Aberration (MDSA) theory described by Viana \textit{et al.}~\cite{Viana2007}. Viana \textit{et al.}'s work includes a description of a beam after propagation through an interface of mismatching refractive indices. An extension of this theory to include two interfaces of mismatched refractive indices, accurately describing the geometry of our experiment, is now presented.

We will first describe the tight focussing of a beam using the integral representation developed by Richards and Wolf~\citep{Wolf1959,Richards1959} which can be extended to include propagation through stratified media~\cite{Torok1997}. Secondly we will give a brief overview of Mie's classic solution that describes light scattering from spheres and finally we will describe how optical forces can be calculated through combining these ideas.

\subsection{Focussed beam description}\label{sec:Richards and Wolf theory}
A Gaussian laser beam with plane wavefronts entering the back aperture of a lens is described by
\begin{equation}
	\boldsymbol{E}_{\text{obj}}(\rho,z) = E_{\text{obj}}e^{ik_{0}z}e^{-\rho^{2}/w^{2}}\boldsymbol{\hat{\epsilon}},
\end{equation}
where $k_{0}=2\pi/\lambda_{0}$, $z$ is the axial direction, $\rho$ is the lateral direction, and $\boldsymbol{\hat{\epsilon}}$ is a unit vector along the wave propagation direction.

When focussed into a medium of refractive index $n_{g}$, in our case glass, the beam will occupy a conical region in space governed by the angle of convergence $\theta_{0}$, and the azimuthal angle $\varphi$. This beam can be thought of as a superposition of plane waves and given by an integral representation of electromagnetic diffraction described by Richards and Wolf~\citep{Richards1959}. The electric field in glass is therefore
\begin{equation}
	\boldsymbol{E}_{\text{glass}} = E_{0}\int_{0}^{2\pi}\int_{0}^{\theta_{0}}\sin{\theta_{g}}\sqrt{\cos{\theta_{g}}}e^{-\gamma^{2}\sin^2{\theta_{g}}}e^{-i\boldsymbol{k\cdot r}_{f}}e^{i\boldsymbol{k\cdot r}}\boldsymbol{\hat{\epsilon}}'(\theta,\varphi)d\theta d\varphi,
	\label{eq:fieldinglass}
\end{equation}
where
\begin{equation}
	E_{0}=-i\frac{n_{g}f}{\lambda_{0}}T_{\text{obj}}E_{\text{obj}},
\end{equation}
and $\theta_{0}$ is the opening angle of the focussed beam given by the NA of a lens, $\theta_{0}=\arcsin{\left(\text{NA}/n_{g}\right)}$, $n_{g}$ is the refractive index of glass, $\theta_{g}$ is the half-cone angle in glass, $T_{obj}$ is the transmission of the objective, $E_{obj}$ is the electric field magnitude at the objective lens back aperture, $\gamma=f/w$ and $f$, $\omega$ and $r_{f}$ are defined in fig.~\ref{fig:Wolf}. $\boldsymbol{\hat{\epsilon}}'(\theta,\varphi)$ rotates the plane waves to occupy all angles from 0 to $\theta_{0}$ and all $\varphi$ using a rotation by Euler angles~\citep{Torok1995,Torok1997} $\left(\varphi,\theta_{g},-\varphi\right)$. From here the NA of a lens describes the opening angle of the cone of focus exterior to and immediately before any interfaces. When the opening angle of the converging beam is larger than the critical angle for the glass to air refractive indices, the NA of the beam is effectively reduced to $\theta_{0}=\arcsin{\left(N_{1}N_{2}\right)}$.

Equation 4 is the classic representation of a beam focussed to a point, $\boldsymbol{r}_{f}$, however, our system differs as it has two interfaces between exit from the lens and reaching the focal point as shown in fig.~\ref{fig:Wolf}. The plane wave components of equation~\ref{eq:fieldinglass} each refract at the interfaces at $z_{1}=-h$ and $z_{2}=-h+\Delta h$. From Snell's law the angle of refraction in the water layer $\theta_{w}=\arcsin\left({\sin{\theta_{g}}/N_{1}}\right)$ and the angle of refraction in air $\theta_{a}=\arcsin\left({\sin{\theta_{w}}/N_{2}}\right)$ where $N_{1}=n_{w}/n_{g}$ and $N_{2}=n_{a}/n_{w}$ are the relative refractive indices of the glass-water and water-air interfaces respectively.

Our beam representation must therefore include the effects of propagation through media of stratified refractive index~\citep{Torok1997}. The focussed beam in the third medium, air, is described by
\begin{equation}
	\boldsymbol{E}_{\text{air}} = E_{0}\int_{0}^{2\pi}\int_{0}^{\theta_{0}}T(\theta_{g})\sin{\theta_{g}}\sqrt{\cos{\theta_{g}}}e^{-\gamma^{2}\sin^2{\theta_{g}}}e^{-i\left(k_{gz}-k_{wz}\right)h}e^{-i\boldsymbol{k\cdot r}_{f}}e^{i\boldsymbol{k_{\textit{a}}\cdot r}}\boldsymbol{\hat{\epsilon}}'(\theta_{a},\varphi_{a})d\theta d\varphi,
	\label{eq:fieldinair}
\end{equation}
where $\left(k_{gz}-k_{wz}\right)h$ takes into account beam propagation in the glass slide up to the first interface, $k_{a}=n_{a}k_{0}$, and each plane wave amplitude is multiplied by its respective Fresnel transmission coefficient
\begin{equation}
T(\theta_{g})=T_{1}(\theta_{g})T_{2}(\theta_{g})=\frac{2\cos{\theta_{g}}}{\cos{\theta_{g}}+N_{1}\cos{\theta_{w}}}\frac{2\cos{\theta_{w}}}{\cos{\theta_{w}}+N_{2}\cos{\theta_{a}}}.
	\label{eq:stratfield}
\end{equation}

The effect of the additional factors in equation~\ref{eq:fieldinair} over equation~\ref{eq:fieldinglass} is to introduce a spherical aberration that deforms the wavefront preventing diffraction limited focussing to the point $\boldsymbol{r}_{f}$. This is quantified in terms of an aberration function as will be shown later in section~\ref{sec:forcetheory}.

Having focussed the beam through two mismatched refractive index interfaces, the height of the paraxial focal plane above the water layer is found from the objective displacement, $X$, through

\begin{equation}
	L = \left(X\frac{n_{w}}{n_{g}}-\Delta h\right)\frac{n_{a}}{n_{w}}.
\end{equation}

Using the work of T\"{o}r\"{o}k and Varga~\citep{Torok1997} and equation~\ref{eq:fieldinair} we calculate the profiles of beams focussed in our system and compare them to the ideal beam assumed in most cases, giving some insight into the physics.

\begin{figure}[!ht]
	\begin{center}
	\subfigure[]{\includegraphics[width=2.5cm]{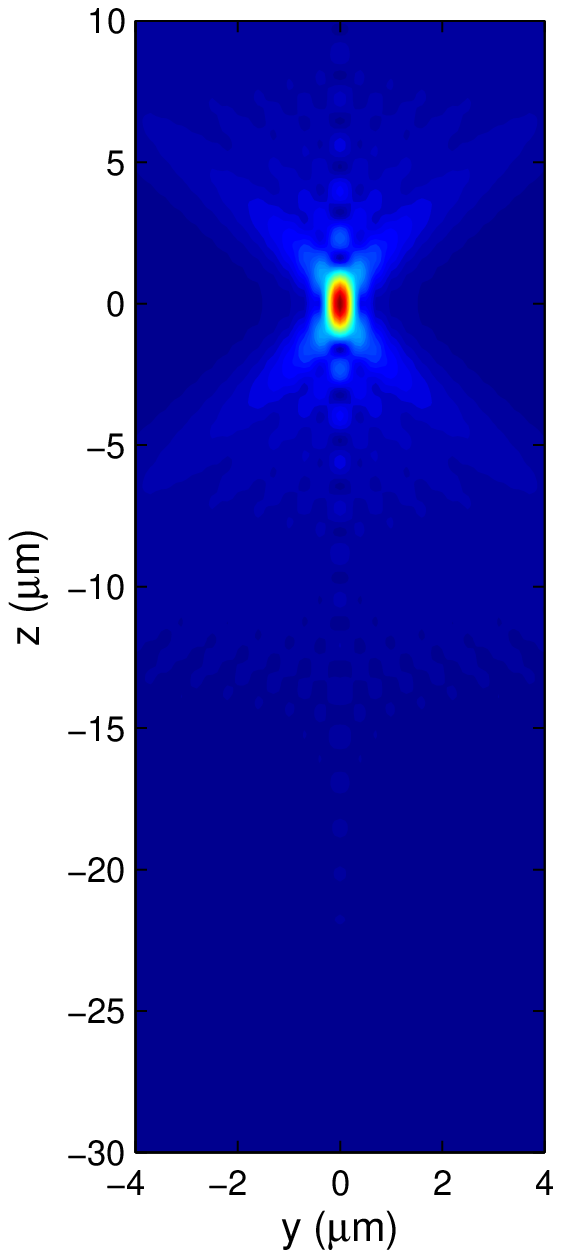}\label{fig:1a}}
	\subfigure[]{\includegraphics[width=2.5cm]{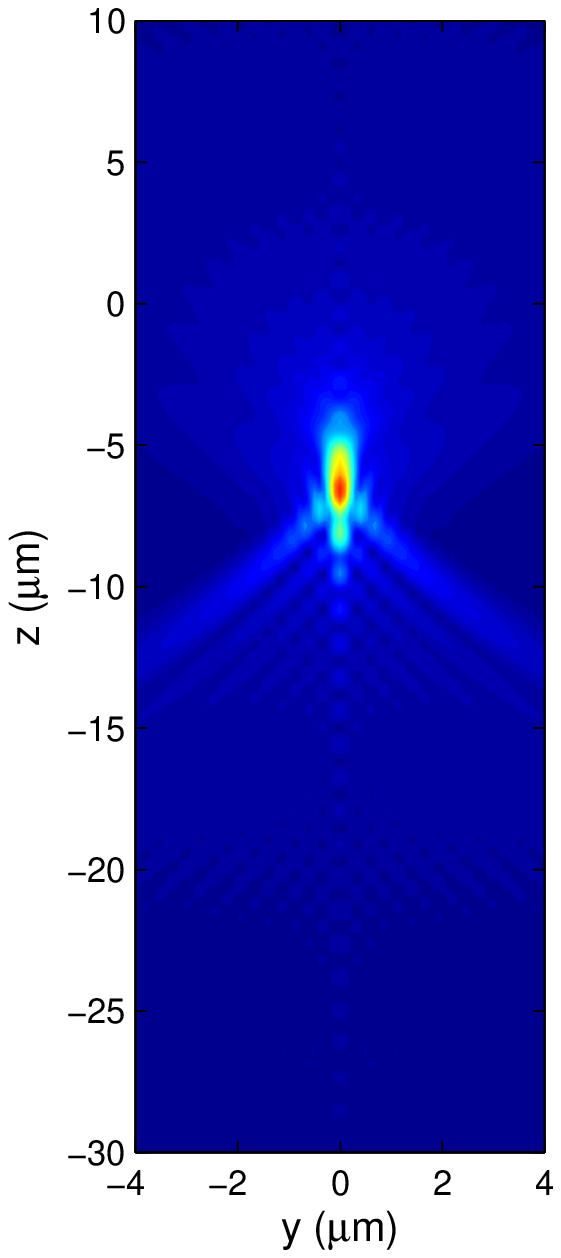}\label{fig:1b}}
	\subfigure[]{\includegraphics[width=2.5cm]{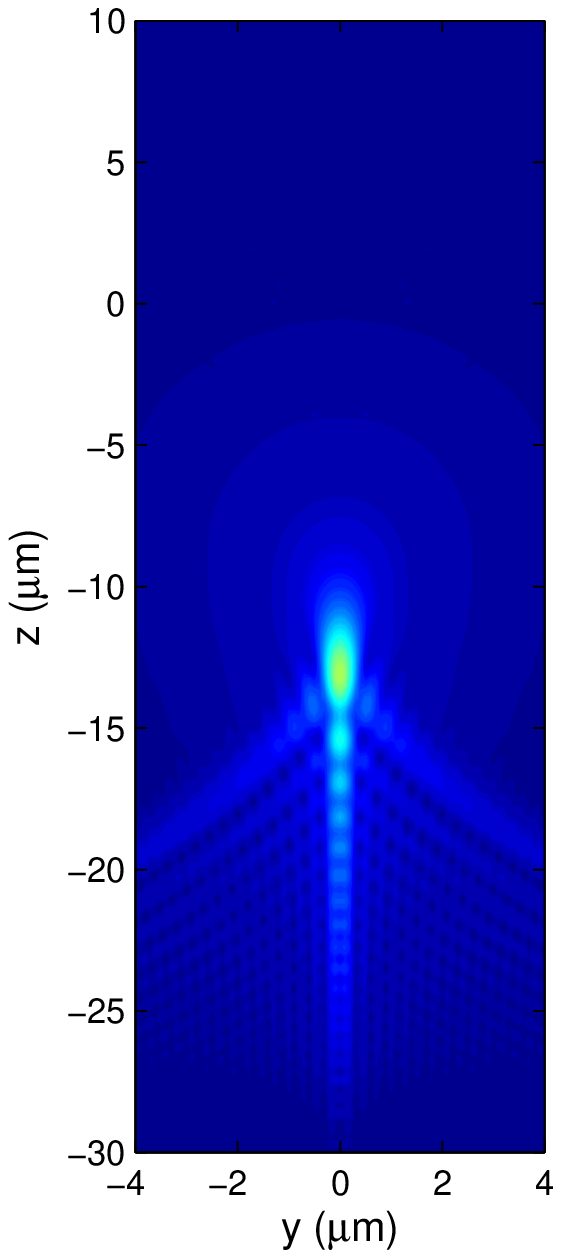}\label{fig:1c}}
	\subfigure[]{\includegraphics[height=5.729cm]{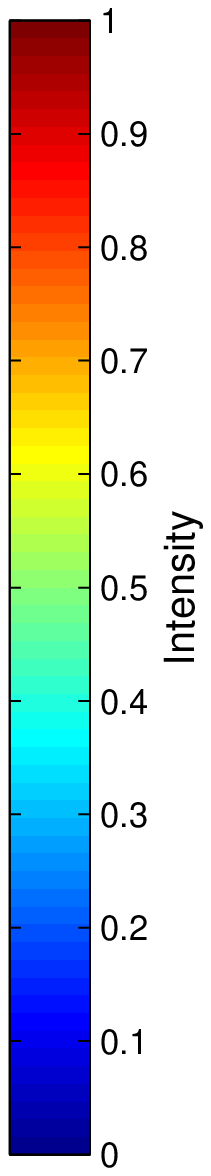}\label{fig:1d}}
	\caption{(Color online) Profile of a focussed $532~\text{nm}$ Gaussian beamstaken from a y-z slice through the beam axis. (a) The beam is focussed into water ($n_{w}=1.33$). (b) The beam is focussed into water having crossed a glass ($n_{g}=1.517$) to water ($n_{w}=1.33$) interface after the lens. (c) The beam is focussed into air ($n_{a}=1.00$) across glass ($n_{g}=1.517$) to water ($n_{w}=1.342$) and water to air interfaces. The objective displacement $X=40~\mu\text{m}$, the water layer is $10~\mu\text{m}$ thick, $\gamma=1$ and $\theta_{0}=41.23^{\circ}$. Zero on the axial axis is the position of the paraxial focus had there been no interfaces.}
	\label{fig:profiles}
	\end{center}
\end{figure}

Figure~\ref{fig:profiles} displays the $yz$-plane beam profiles for focussing in water, through a glass-water interface, and through glass-aqueous-air interfaces. The axial displacement zero point is the position that the paraxial focus, $\boldsymbol{r}_{f}$, would exist at when no refractive index interfaces are present.

The beam focussed in water with no preceding interfaces varies smoothly at the focus compared to those focussed in water and air having first travelled through glass coverslips. In particular the beam focussed to a point in air has a large number of oscillations in intensity along the beam axis. Previous work has shown such landscapes can interact with particles in a non-trivial manner~\citep{Milne2007a}. Particles that are relatively large may not `see' the oscillations while small particles could be trapped at more than one of the `hot-spots'. The colour scale remains the same for all plots so the maximum intensity is less in an airborne trap than for others given the same input power.

Having described beam focussing more realistically, specifically for stratified media, we will move onto the scattering of light by the particles we wish to model in these beams.

\subsection{Mie scattering}\label{sec:MieTheory}
A full and clear derivation of Mie theory can be found in Bohren and Huffman~\citep{Bohren1983} so here we only include the results that are important for our work.

A plane wave incident on a spherical particle results in a scattered electric field that is dependent on the Mie scattering coefficients $a_{n}$ and $b_{n}$~\citep{Bohren1983} where $n$ appears because of the Legendre polynomials in the solution which have $n$ degrees. To simplify the following the Riccati-Bessel functions are introduced as;
\begin{equation}
	\psi_{n}(kr_{s}\xi)=kr_{s}J_{n}(kr_{s})\quad\text{and}\quad\xi_{n}(kr_{s})=kr_{s}h_{n}^{(1)}(kr_{s}),
\end{equation}
where $k$ is the wavenumber, $r_{s}$ is from the spherical coordinates system and $h_{n}^{(1)} = J_{n}+iy_{n}$ is the spherical Hankel function with the spherical Bessel functions $J_{n}$ and $y_{n}$. Using these the scattering coefficients are
\begin{equation}
	a_{n}=\frac{m\psi_{n}(m\xi)\psi_{n}^{'}(\xi)-\psi_{n}(\xi)\psi_{n}^{'}(m\xi)}{m\psi_{n}(m\xi)\xi_{n}^{'}(\xi)-\xi_{n}(\xi)\psi_{n}^{'}(m\xi)}
\end{equation}
and
\begin{equation}
	b_{n}=\frac{\psi_{n}(m\xi)\psi_{n}^{'}(\xi)-m\psi_{n}(\xi)\psi_{n}^{'}(m\xi)}{\psi_{n}(m\xi)\xi_{n}^{'}(\xi)-m\xi_{n}(\xi)\psi_{n}^{'}(m\xi)}
\end{equation}
where $\xi=n_{m}k_{0}R$ is the size parameter and $m=n_{p}/n_{m}$, the relative refractive index of particle to medium.

The previous sections give the necessary background to now allow a description of how the force in optical traps is calculated.

\subsection{Force calculation}\label{sec:forcetheory}
To calculate the force, $F$, we follow the full electromagnetic approach, integrating the Maxwell stress tensor over the surface of the object;

\begin{equation}
	\left\langle F\right\rangle = \left\langle\oint_{S}\boldsymbol{\hat{n}}\cdot\boldsymbol{T}dS\right\rangle,
\end{equation}
where $\boldsymbol{\hat{n}}$ is the surface normal and $\boldsymbol{T}$ is the electromagnetic stress tensor. Due to system symmetry and also momentum conservation the force simplifies so the surface is at infinity, thus giving

\begin{equation}
	F = \lim_{r\rightarrow\infty}\left(-\frac{r}{2}\int_{S_{r}}\boldsymbol{r}\left(\epsilon E^{2} + \mu_{0}H^{2}\right)\right),
\end{equation}
where $\boldsymbol{E} = \boldsymbol{E}_{inc}+\boldsymbol{E}_{scat}$, with equivalents for the magnetic field~\citep{Mazolli2003}.

An analytical solution to this can be complicated, even for spheres, as seen in Barton \textit{et al.}~\citep{Barton1989}. We take the simpler approach described by Mazolli \textit{et al.}~\citep{Mazolli2003} where the vector electric and magnetic fields are given in terms of scalar Debye potentials, also known as Hertz vectors~\citep{Born1980}. The optical forces are calculated by following Farsund and Felderhof~\citep{Farsund1996}, who derive force, torque and absorbed energy for an object of arbitrary shape and material given the Debye potentials for the incident and scattered fields. The analytical solutions for force found through Farsund and Felderhof~\citep{Farsund1996} are converted to trapping efficiency through equation~\ref{eq:momentumefficiency}.

Debye~\cite{Debye1909} calculated the force on a sphere due to an incident plane wave, and here the result is generalised to a focussed beam. First, the Debye potential for a single plane wave is~\citep{Mazolli2003}
\begin{equation}
	\Pi_{\boldsymbol{k}(r,\theta,\varphi)}^{E} = \frac{E_{0}}{k}\sum_{j=1}^{\infty}i^{j-1}J_{j}(kr)\sqrt{\frac{4\pi\left(2j+1\right)}{j\left(j+1\right)}}\sum_{m=-j}^{j}e^{-i\left(m-1\right)\varphi_{k}}d_{m,1}^{j}(\theta_{k})Y_{jm}(\theta,\varphi).
\end{equation}
where $J_{j}$ are the spherical Bessel functions and $Y_{jm}$ are the spherical harmonics. Using the matrix elements of finite rotations, also known as Wigner $d$ functions~\citep{Blanco1997,Edmonds1957}, for rotation in the basis of spherical harmonics, the Debye potential for a focussed Gaussian beam made from a superposition of plane waves, whose field is represented by equation~\ref{eq:fieldinair}, is
\begin{multline}
	\Pi_{inc}^{E}(r,\theta,\varphi) = \frac{E_{0}}{k}\int_{0}^{\theta_{0}}\sin{\theta_{k}}\sqrt{\cos{\theta_{k}}}e^{-\gamma^{2}\sin^2{\theta_{k}}}\sum_{j=1}^{\infty}i^{j-1}J_{j}(kr)\sqrt{\frac{4\pi\left(2j+1\right)}{j\left(j+1\right)}}\\\times\sum_{M=-j}^{j}d_{m,1}^{j}(\theta_{k})Y_{jm}(\theta,\varphi)\int_{0}^{2\pi}e^{-i\boldsymbol{k\cdot r_{\text{f}}}}e^{-i\left(k_{gz}-k_{wz}\right)h}e^{-i\left(m-1\right)\varphi_{k}}d\varphi_{k}.
\end{multline}
Evaluating the integral over the azimuthal angle the Debye potential for the incident field becomes
\begin{multline}
	\Pi_{inc}^{E}(r,\theta,\varphi) = \frac{E_{0}}{k}\int_{0}^{\theta_{0}}\sin{\theta_{k}}\sqrt{\cos{\theta_{k}}}e^{-\gamma^{2}\sin^2{\theta_{k}}}\sum_{j=1}^{\infty}i^{j-1}J_{j}(kr)\sqrt{\frac{4\pi\left(2j+1\right)}{j\left(j+1\right)}}\\\times\sum_{m=-j}^{j}d_{m,1}^{j}(\theta_{k})Y_{jm}(\theta,\varphi)2\pi\left(-i\right)^{m-1}e^{-ikz_{f}\cos{\theta_{k}}}e^{-i\left(k_{gz}-k_{wz}\right)h}\\\times J_{m-1}(k\rho_{R}\sin{\theta_{k}})e^{-\left(m-1\right)\varphi_{f}}.
	\label{eq:Debyeincidnet}
\end{multline}
From fig.~\ref{fig:Wolf} the relative locations of the planes gives
\begin{equation}
	z_{f} = \frac{1}{N_{1}}\left(\Delta h + \frac{L_{2}}{N_{2}}\right) - L_{1} - z.
\end{equation}
Substituting this into the middle two exponents of equation~\ref{eq:Debyeincidnet} we derive the aberration function~\citep{Dutra2007,Viana2007,Torok1995,Torok1997}, $\Psi$, of our system to be
\begin{equation}
	\Psi = k_{0}\left(-\left(\frac{n_{g}}{N_{1}}\Delta{h}+\frac{n_{g}}{N_{1}N_{2}}L_{2}\right)\cos{\theta_{g}} + n_{w}\Delta h\cos{\theta_{w}} + n_{a}\left(L_{2} + z\right)\cos{\theta_{a}}\right).
\end{equation}

The Debye potential for the scattered field is found through the incident fields interaction with a sphere and hence is dependent on Mie coefficient $a_{j}$ and Hankel function $h_{j}^{(1)}$ such that
\begin{multline}
	\Pi_{inc}^{E}(r,\theta,\varphi) = -2\pi\frac{E_{0}}{k}\sum_{j=1}^{\infty}\sum_{m=-j}^{j}i^{j-m}G_{jm}(\rho_{f},z_{f})e^{-\left(m-1\right)\varphi_{f}}\sqrt{\frac{4\pi\left(2j+1\right)}{j\left(j+1\right)}}\\\times a_{j}h_{j}^{(1)}(kr)Y_{jm}(\theta,\varphi),
	\label{eq:Debyescattered}
\end{multline}
where
\begin{equation}
	G_{jm}=\int_{0}^{\theta_{0}}T(\theta)\sin{\theta}\sqrt{\cos{\theta}}e^{-\gamma^{2}\sin^2{\theta}}d_{m,1}^{j}(\theta_{a})J_{m-1}(k\rho\sin{\theta_{a}})e^{i\Psi(z,\theta)}d\theta.
\end{equation}
Similar expressions can be found for the magnetic field, $\boldsymbol{H}$, using the Mie coefficient $b_{n}$.

The efficiencies are given for the lateral and axial components each with two separate contributions, one for the rate of removal of momentum from the incident beam, $Q_{e}$, and the other for minus the rate of momentum transfer to the scattered field, $Q_{s}$, so the total efficiency $Q_{tot}^{\rho,z} = Q_{s}^{\rho,z} + Q_{e}^{\rho,z}$. The forces are calculated for circularly polarised light but can equally but done for linear polarisations~\citep{Dutra2007}. The axial component of the trapping efficiency is given by~\citep{Viana2007,Mazolli2003,Neto2000,Dutra2007};
\begin{equation}
Q_{e}^{z}=\frac{4\gamma^{2}}{AN_{1}N_{2}}\mathcal{R}\sum_{j=1}^{\infty}\sum_{m=-j}^{j}\left(2j+1\right)\left(a_{j}+b_{j}\right)G_{j,m}G_{j,m}^{'\ast}
	\label{eq:Qz}
\end{equation}
and
\begin{equation}\begin{split}
	Q_{s}^{z}=\frac{8\gamma^{2}}{AN_{1}N_{2}}\mathcal{R}\sum_{j=1}^{\infty}\sum_{m=-j}^{j}\left(\frac{\sqrt{j\left(j+2\right)\left(j-m+1\right)\left(j+m+1\right)}}{j+1}\left(a_{j}a_{j+1}^{\ast}+b_{j}b_{j+1}^{\ast}\right)\right.\\
\left.\times G_{j,m}G_{j+1,m}^{\ast}+\frac{2j+1}{j\left(j+1\right)}ma_{j}b_{j}^{\ast}\left|G_{j,m}\right|^{2}\right).
\end{split}\end{equation}
The lateral efficiencies are
\begin{equation}	Q_{e}^{\rho}=\frac{2\gamma^{2}}{AN_{1}N_{2}}\mathcal{I}\sum_{j=1}^{\infty}\sum_{m=-j}^{j}\left(2j+1\right)\left(a_{j}+b_{j}\right)G_{j,m}\left(G_{j,m+1}^{-}-G_{j,m-1}^{+}\right)^{\ast}
\end{equation}
and
\begin{multline}
Q_{s}^{\rho}=\frac{8\gamma^{2}}{AN_{1}N_{2}}\mathcal{I}\sum_{j=1}^{\infty}\sum_{m=-j}^{j}\frac{\sqrt{j\left(j+2\right)\left(j-m+1\right)\left(j+m+1\right)}}{j+1}\left(a_{j}a_{j+1}^{\ast}+b_{j}b_{j+1}^{\ast}\right)\\\times\left(G_{j,m}G_{j+1,m+1}^{\ast}+G_{j,-m}G_{j+1,-m-1}^{\ast}\right),
	\label{eq:Qrho}
\end{multline}
where $A$ is the fraction of the beam power that enters the objective back aperture, to account for overfilling~\citep{Viana2007}, and the functions $G_{j,m}^{'}$ and $G_{j,m}^{\pm}$ are defined as
\begin{equation}
G_{jm}^{'}=\int_{0}^{\theta_{0}}T(\theta)\sin{\theta}\sqrt{\cos{\theta}}\cos{\theta_{a}}e^{-\gamma^{2}\sin^2{\theta}}d_{m,1}^{j}(\theta_{a})J_{m-1}(k\rho\sin{\theta_{a}})e^{i\Psi(z,\theta)}d\theta.
\end{equation}
and
\begin{equation}
G_{jm}^{\pm}=\int_{0}^{\theta_{0}}T(\theta)\sin{\theta}\sqrt{\cos{\theta}}\sin{\theta_{a}}e^{-\gamma^{2}\sin^2{\theta}}d_{m\pm1,1}^{j}(\theta_{a})J_{m-1}(k\rho\sin{\theta_{a}})e^{i\Psi(z,\theta)}d\theta.
\end{equation}
In the limiting case where $\Delta h=0$ and $n_{w}=n_{g}$ the results return to those of Viana \textit{et al.}~\citep{Viana2007} for a glass to water interface without an intermediate aqueous layer. For $\Delta h=0$ and $n_{g}=n_{a}=n_{w}$ and $X=0$, the results of Mazolli \textit{et al.}~\citep{Mazolli2003}, and Neto and Nussenzveig~\citep{Neto2000} are matched.

The final crucial point concerns the computation of these equations. Rather than completing the sums in equations~\ref{eq:Qz} to~\ref{eq:Qrho} \textit{ad infinitum} it is useful to know that it is sufficient to sum over $j$ up to $\xi_{s} + 4\xi_{s}^{1/3} + 2$, or its nearest integer, due to the convergence of the Mie scattering coefficients (Appendix A in Bohren and Huffman~\citep{Bohren1983}).

Having introduced the theory and visualised the focussed beams we will move onto examining the outcome of applying the theories to `normal' optical traps and then to our application.

\section{Results and discussion}
Ashkin observed that even with relatively loose focussing of a Gaussian beam, particles (with $m>1$) always had the tendency to move toward the beam centre where they would reach a lateral equilibrium position. In airborne tweezing this occurs in exactly the same manner so its modelling is not of great importance. It is the axial efficiency and force curves, and the associated balance between gradient and scattering forces, that governs whether a particle is trapped or not. Although for completeness we have given the analytical solutions for lateral efficiencies, it will be the axial direction we consider as it determines the unusual phenomena observed.

The results we present should be considered a set of typical examples that can be produced using our code and is by no means exhaustive. We also note that we have included several results pertaining to the optical trapping of spheres in water to highlight the large difference between those experiments and those in air.

\subsection{Comparison of geometrical optics and Mie scattering}
\begin{figure}[!ht]
	\begin{center}
	\subfigure[]{\includegraphics[width=8.6cm]{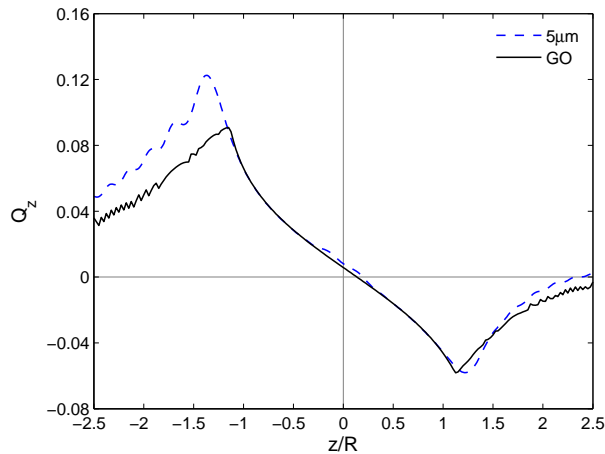}}
	\\
	\subfigure[]{\includegraphics[width=8.6cm]{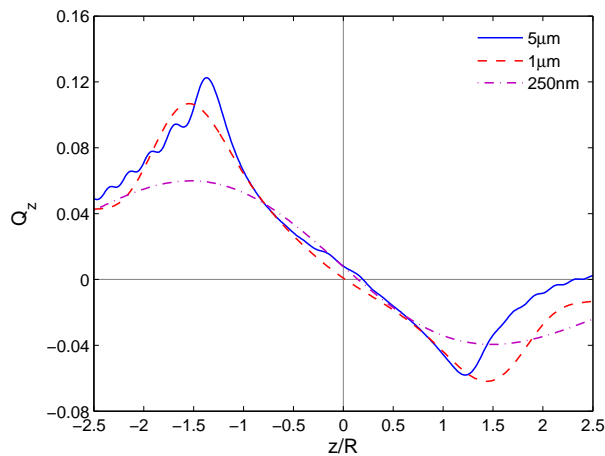}}
	\caption{(Color online) Axial trapping efficiency calculated through GO and Mie theories when $250~\text{nm}$, $1~\mu\text{m}$, and $5~\mu\text{m}$ silica spheres ($n_{p}=1.445$) are trapped with $532~\text{nm}$ light in a water medium ($n_{m}=1.33$) with $\gamma=1$ and $\theta_{0}=61.25^{\circ}$ in a system like fig.~\ref{fig:Wolf} with no refractive index interfaces. The four curves are plotted on two separate graphs for clarity. In (a) the black solid line is calculated through GO and the blue dashed line is calculated through Mie scattering. In (b) all curves are calculated with Mie scattering.}
	\label{fig:GOMie}
	\end{center}
\end{figure}

First we will make a comparison between the theoretical predictions of GO against those from Mie scattering. In this first instance we will neglect the effects of spherical aberration and show in fig.~\ref{fig:GOMie} the axial trapping efficiency calculated through both theories when a $250~\text{nm}$, $1~\mu\text{m}$, and $5~\mu\text{m}$ silica sphere is trapped with $532~\text{nm}$ light in water. As described in the theory section there is a limit to the opening angle of the focussed light and hence NA of the trapping beam. For a beam focussed through a coverslip-water interface this limit is $\theta_{0}=\theta_{c}\simeq62^{\circ}$, thus $\textnormal{NA} = 1.33\sin{\theta_{c}}=1.17$, and is the value used for this first test.

Clearly the prediction of GO disagrees with those of Mie scattering. However, GO stands up surprisingly well even for spheres with radii similar to the wavelength of trapping light. Testing the theory on a $5~\mu\text{m}$ sphere, which is approaching the regime where GO should become applicable, it is indeed a reasonable approximation except for the area closest to the paraxial focus and at the extremities.

The inaccuracies are unsurprising considering the earlier discussion of the limits of GO. For small spheres Mie scattering plays a dominant role that differs to simple ray optics and for the larger sphere the non-Gaussian beam focus plays the important role that GO cannot account for.

\begin{figure}[!ht]
	\begin{center}
	\subfigure[]{\includegraphics[width=8.6cm]{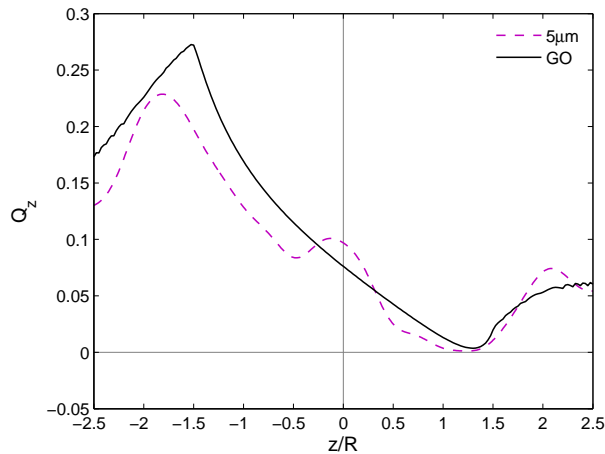}}
	\\
	\subfigure[]{\includegraphics[width=8.6cm]{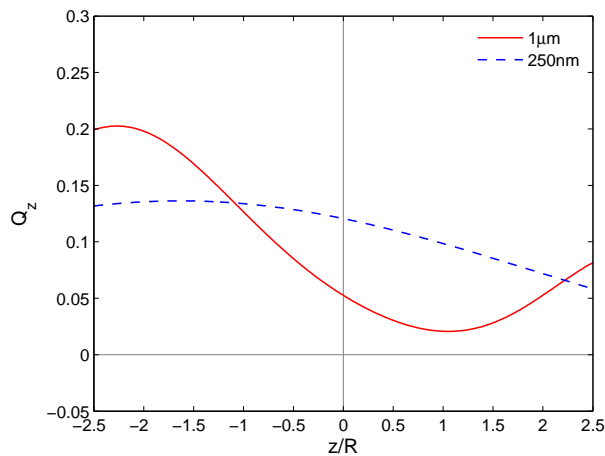}}
	\caption{(Color online) Axial trapping efficiency calculated through GO and Mie theories for $250~\text{nm}$, $1~\mu\text{m}$, and $5~\mu\text{m}$ water droplets ($n_{p}=1.342$) trapped with $532~\text{nm}$ light in air ($n_{m}=1.000$) with $\gamma=1$ and $\theta_{0}=41.23^{\circ}$ in a system like fig.~\ref{fig:Wolf} with no refractive index interfaces. The four curves are plotted on two separate graphs for clarity. In (a) the black solid line is calculated through GO and the purple dashed line is calculated through Mie theory. In (b) both curves are calculated with Mie theory.}
	\label{fig:aircontrast}
	\end{center}
\end{figure}

We will now test how applicable GO is when trapping objects in air. Here the upper limit of the NA is reduced ($\theta_{0}=\theta_{c}\simeq41.2^{\circ}$ therefore $NA\simeq0.66$) and lends itself toward the paraxial approximation, hence GO. Yet the ratio of particle to medium refractive index is higher than in colloidal systems thus moving further into the applicability of Mie scattering. Figure~\ref{fig:aircontrast} plots the axial efficiency for the same particles as fig.~\ref{fig:GOMie} except the medium is now air ($n_{m}=1.00$) and the particle is a water droplet ($n_{p}=1.342$).

The form of the GO curve remains similar to the previous example except with an absolute increase in efficiency. This is probably due to an increased scattering force component from the larger particle-medium refractive index contrast and reduction in NA. GO predicts the droplet will `just' not obtain an axial equilibrium position allowing it a brief reprieve in matching the more rigorous Mie theory. However, this disappears quite quickly when noting the drastic curve change as the forces on three sizes of spheres are computed using Mie scattering. The largest sphere, $5~\mu\text{m}$, enters the beginning of the GO regime ($R\gg\lambda$), yet the theory completely fails to indicate the occurrence of a second minima, predicted by Mie theory.

It has been shown here that GO, although not highly accurate, can provide reasonable predictions of the efficiencies of trapping colloidal particles in water, giving indications of what one would expect in real world systems. However, in the same manner it has been shown that GO is not an appropriate description of airborne trapping with the wild variation as a function of radius not predicted, and the inability to predict important features.

Our next extension is where the GO description falls down, namely in the consideration of the relevance of spherical aberration for both colloidal and airborne systems. Only Mie theory can accommodate variations in the phase wavefront profile so now only this will be considered.

\begin{figure}[!ht]
	\begin{center}
	\subfigure[]{\includegraphics[width=8.6cm]{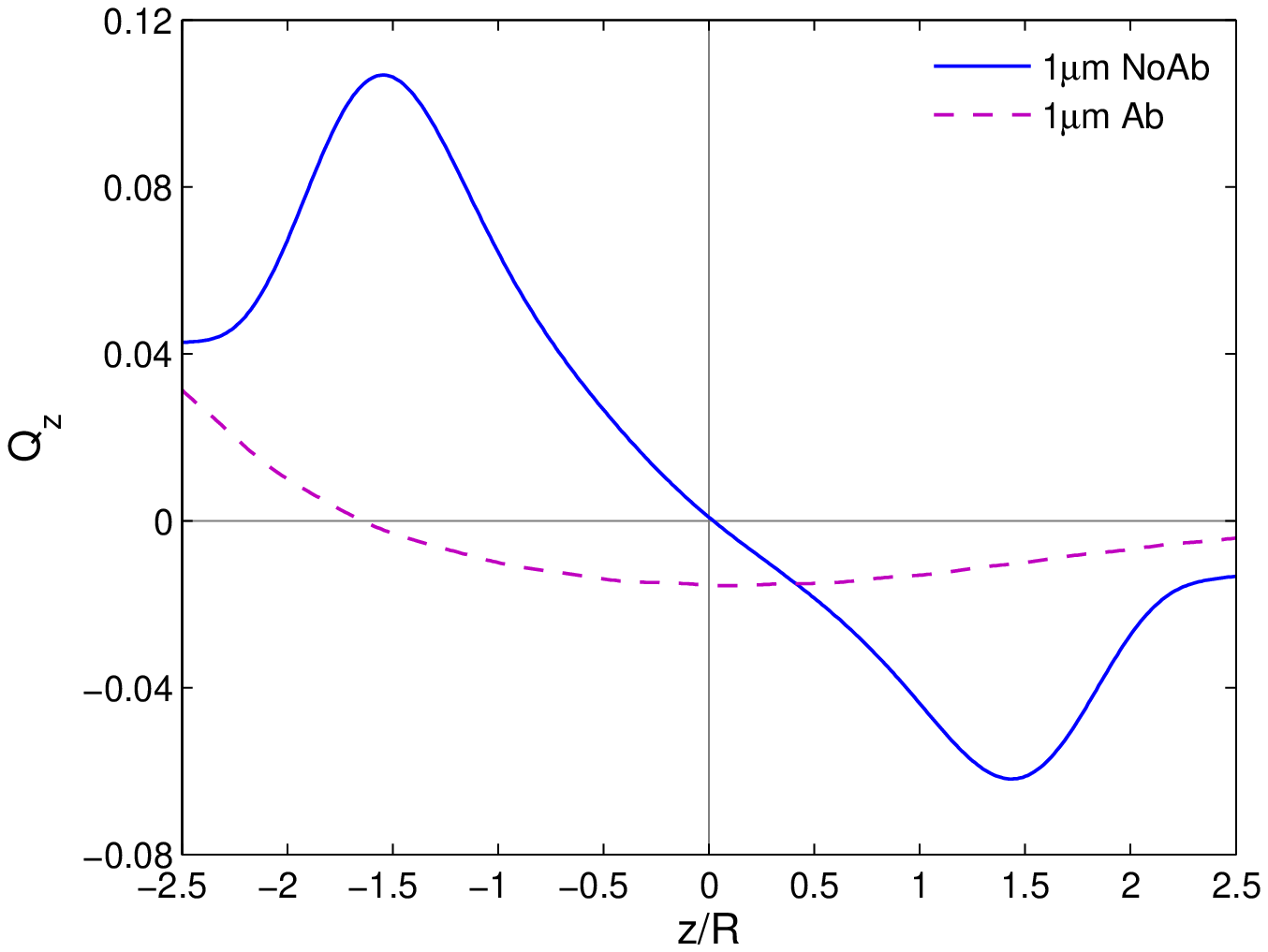}}
	\\
	\subfigure[]{\includegraphics[width=8.6cm]{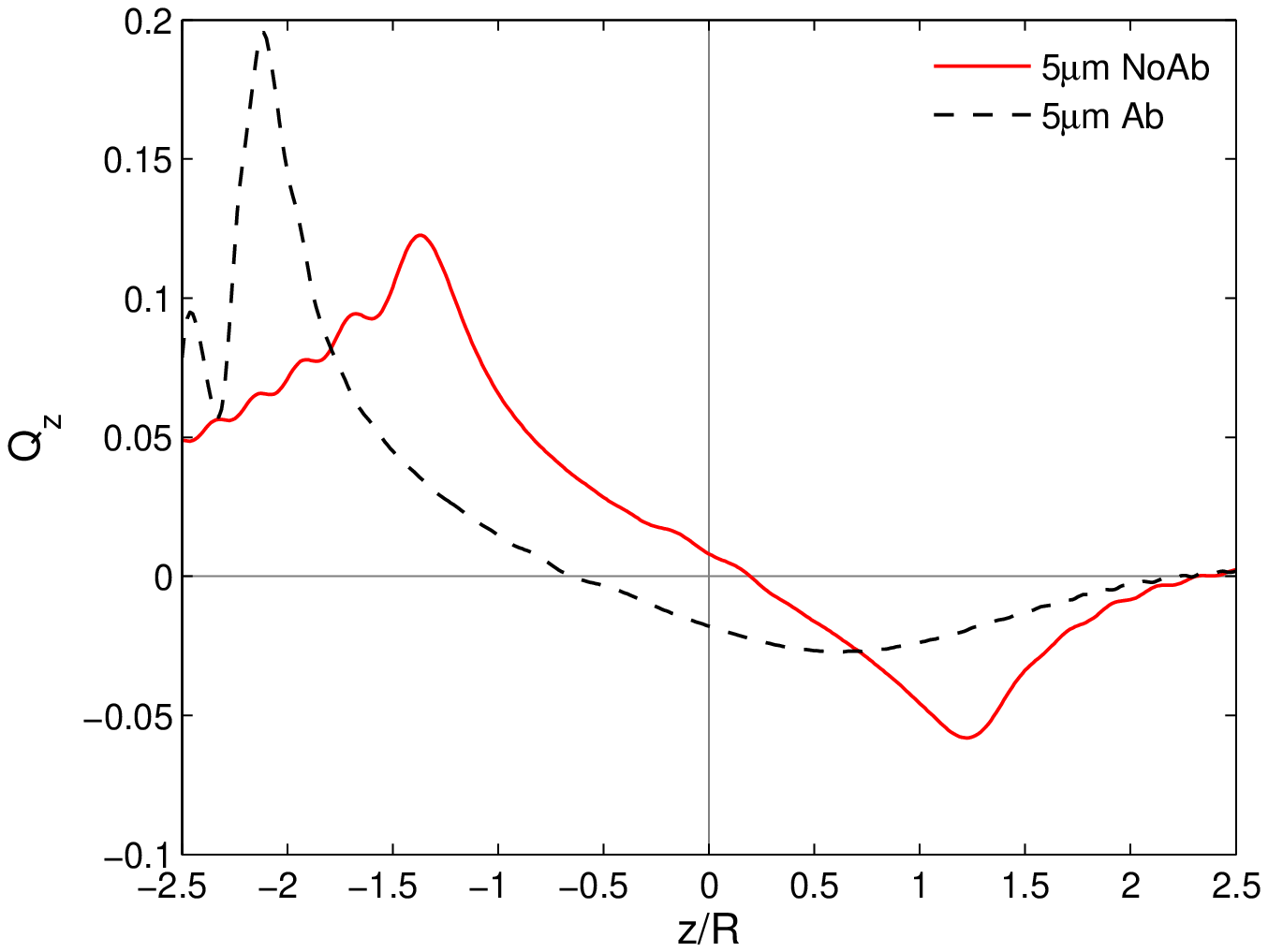}}
	\caption{(Color online) Axial efficiency curves for $1~\mu\text{m}$ and $5~\mu\text{m}$ silica spheres ($n_{p}=1.445$) trapped in water above a glass coverslip with and without the aberration induced by the refractive index interface of glass ($n_{g}=1.517$) to water ($n_{w}=1.33$) taken into account. The objective axial displacement $X = 35~\mu\text{m}$, $\gamma=1$ and $\theta_{0}=61.25^{\circ}$. (a) For a $1~\mu\text{m}$ sphere the blue solid line is without aberration and the purple dashed line with aberration. (b) For a $5~\mu\text{m}$ sphere the red solid line is without aberration and the black dashed line with aberration.}
	\label{fig:colloidSA}
	\end{center}
\end{figure}

Figure~\ref{fig:colloidSA} plots the axial efficiency curves for $1~\mu\text{m}$ and $5~\mu\text{m}$ silica spheres trapped in a water medium when the aberration due to a single coverslip-water interface is and is not neglected.

Spherical aberration clearly has a significant effect on optical trapping efficiency curves as expected~\citep{Fallman2003,Im2003}. There is a drastic difference between the axial efficiencies from aberrated and non-aberrated beams. The two main effects are a reduction in $Q_{z,max}^{-}$, reducing the axial trap strength, and a decrease in axial equilibrium position, $z_{eq}$, so the spheres `sit' lower in the trap relative to the paraxial focal point.

\begin{figure}[!ht]
	\begin{center}
	\subfigure[]{\includegraphics[width=8.6cm]{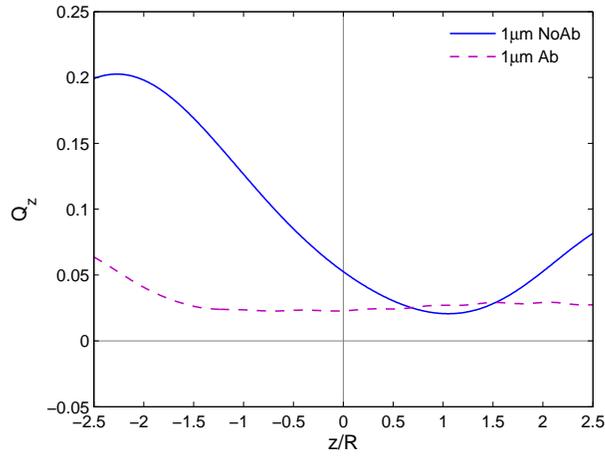}\label{fig:waterabA}}
	\\
	\subfigure[]{\includegraphics[width=8.6cm]{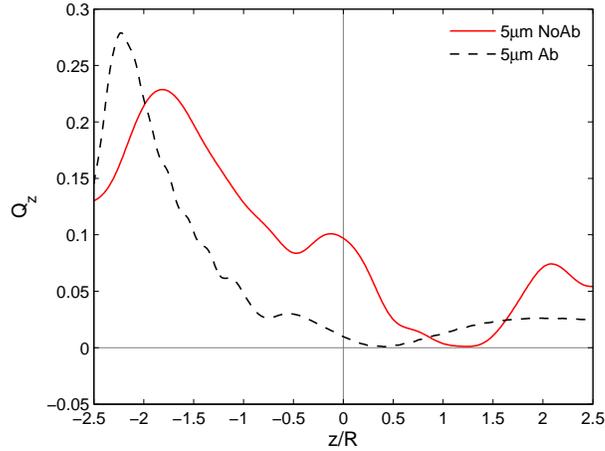}\label{fig:waterabB}}
	\caption{(Color online) Axial efficiency curves for $1~\mu\text{m}$ and $5~\mu\text{m}$ water droplets ($n_{p}=1.342$) trapped in air ($n_{a}=1.000$) above a glass coverslip ($n_{g}=1.517$) and thin water layer ($n_{w}=1.342$) as depicted in fig.~\ref{fig:Wolf}. The objective axial displacement $X=40~\mu\text{m}$, the water layer is $10~\mu\text{m}$ thick, $\gamma=1$ and $\theta_{0}=41.23^{\circ}$. (a) For a $1~\mu\text{m}$ sphere the blue solid line is without aberration and the purple dashed line with aberration. (b) For a $5~\mu\text{m}$ sphere the red solid line is without aberration and the black dashed line with aberration.}
	\label{fig:waterab}
	\end{center}
\end{figure}

Spherical aberration plays a major role in the physics describing optical trapping, so it must be considered in our system (fig.~\ref{fig:Wolf}) where there are two interfaces with mismatched refractive indices. Repeating the previous figure for water droplets trapped in air above a coverslip and thin aqueous layer we obtain fig.~\ref{fig:waterab}.

\begin{figure}[!ht]
	\begin{center}
	\subfigure[]{\includegraphics[width=8.6cm]{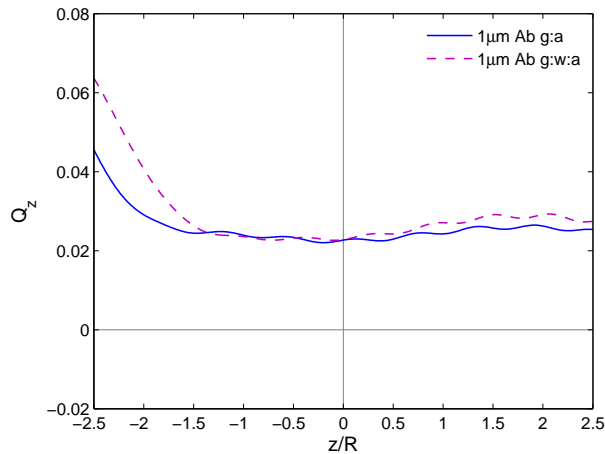}\label{fig:waterab2A}}
	\\
	\subfigure[]{\includegraphics[width=8.6cm]{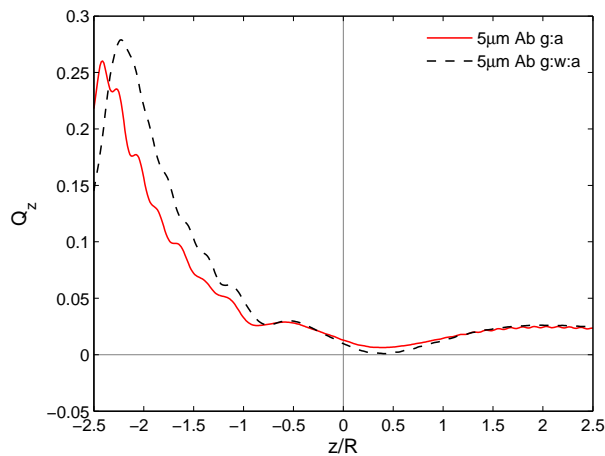}\label{fig:waterab2B}}
	\caption{(Color online) Axial efficiency curves for $1~\mu\text{m}$ and $5~\mu\text{m}$ water droplets ($n_{p}=1.342$) trapped in air ($n_{a}=1.000$) above a glass coverslip ($n_{g}=1.517$) with and without a thin water layer ($n_{w}=1.342$). The objective axial displacement, $X=40~\mu\text{m}$, $\gamma=1$, $\theta_{0}=41.23^{\circ}$ and when the thin water layer exists it is $10~\mu\text{m}$ thick. In (a) the blue solid and purple dashed curves are calculated without and with the thin water layer respectively. In (b) the red solid and black dashed curves are calculated without and with the thin water layer respectively.}
	\label{fig:waterab2}
	\end{center}
\end{figure}

The inclusion of spherical aberration in the description greatly affects the efficiency curves for airborne water droplets. There is a reduction in $Q_{z,max}^{-}$, reducing the axial trap strength and for the larger droplet a general `smoothing' of the curve occurs with smaller local minima created. Figure~\ref{fig:waterab2}, for the same system and objective displacement but with and without the aqueous layer, shows a change in the axial curves indicating the significance of the aqueous layer and its necessary inclusion in the theory.

Having established the most appropriate model to use and the physics to include - primarily the effect of the stratified layers between the objective and trapped particle, spherical aberration and the high relative refractive index - we will now move on to attempt to explain the phenomena observed as described in the introduction.

\subsection{Predicting experimental observations}
\begin{figure}[!ht]
	\begin{center}
	\includegraphics[width=8.6cm]{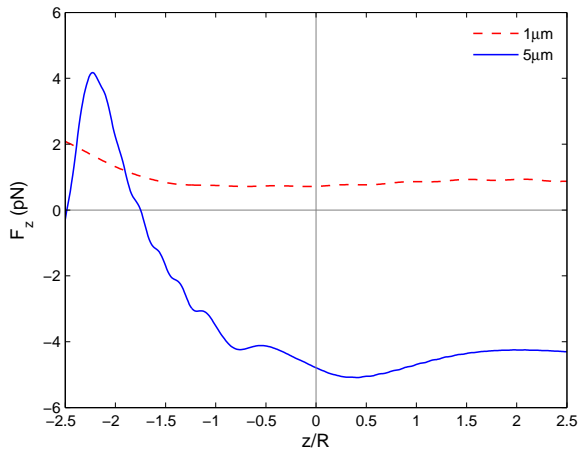}
	\caption{(Color online) Axial force curves for $1~\mu\text{m}$ and $5~\mu\text{m}$ water droplets ($n_{p}=1.342$) trapped in air ($n_{a}=1.000$) above a glass coverslip ($n_{g}=1.517$) and thin water layer ($n_{w}=1.342$). The objective axial displacement, $X=40~\mu\text{m}$, the water layer is $10~\mu\text{m}$ thick, $\gamma=1$, $\theta_{0}=41.23^{\circ}$. The red dashed line is for a $1~\mu\text{m}$ droplet and the blue solid line is for a $5~\mu\text{m}$ droplet.}
	\label{fig:airforce}
	\end{center}
\end{figure}

All the previous theoretical results shown have neglected any particle buoyancy. For colloidal systems this is a reasonable approximation with the density of the trapped objects approximately that of the medium. Thus, there is only a multiplicative factor between efficiency and force graphs via equation~\ref{eq:momentumefficiency} which allows the axial $Q$ curves to be treated as scaled force curves. However, this is a very poor assumption when considering water droplets suspended in air with the large density contrast: $\rho_{water}\simeq1000\rho_{air}$. In order to fully appreciate what the theory predicts we must calculate the force experienced by the microsphere using equation~\ref{eq:momentumefficiency} and subtract its weight. The droplet and system parameters from fig.~\ref{fig:waterab} including aberration is replicated with a trapping power of $10~\text{mW}$ to calculate the corresponding force curves in fig.~\ref{fig:airforce}.

For the $1~\mu\text{m}$, unlike the $5~\mu\text{m}$, droplet the force plot has no significant effect on the properties deduced from the efficiency plot. The weight of the $5~\mu\text{m}$ droplet is comparable to the optical force so that an equilibrium position,  $z_{eq}$, exists which was not indicated in the optical efficiency curve.

\begin{figure}[!ht]
	\begin{center}
	\subfigure[]{\includegraphics[width=8.6cm]{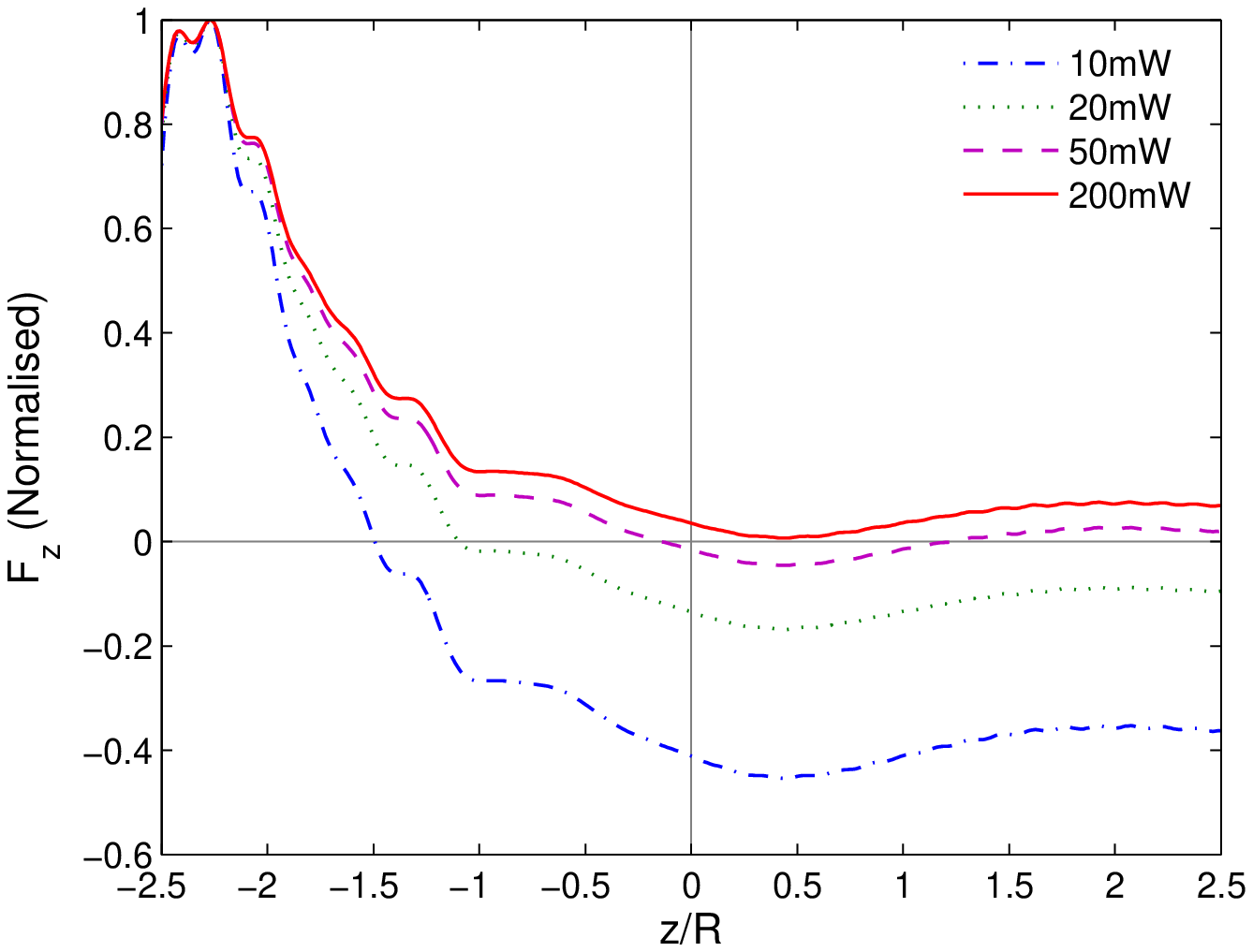}\label{fig:pgA}}
	\\
	\subfigure[]{\includegraphics[width=8.6cm]{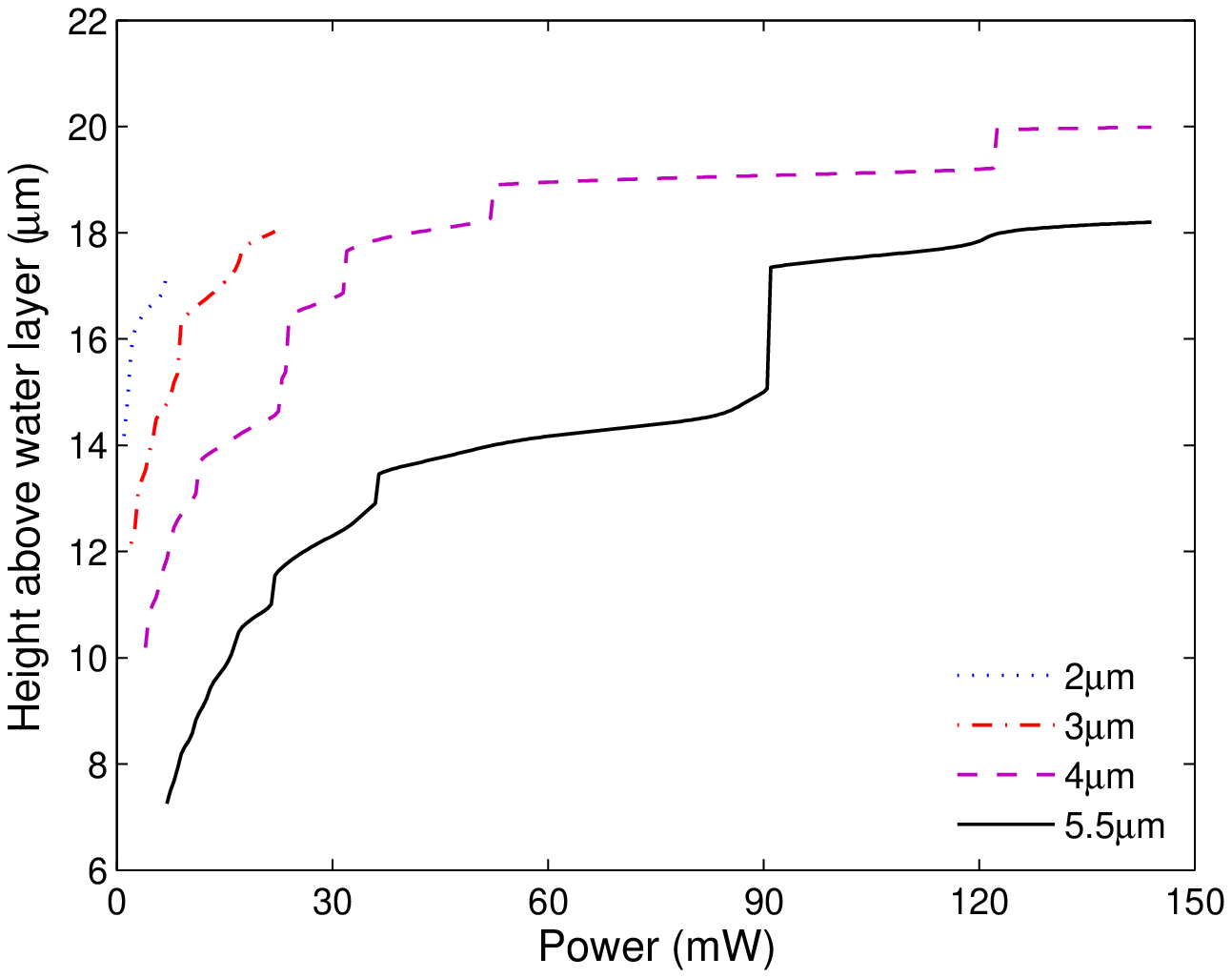}\label{fig:pgB}}
	\caption{(Color online) (a) Variation of axial force for a $4~\mu\text{m}$ water droplet ($n_{p}=1.342$) trapped in air ($n_{a}=1.000$) at trapping powers of $10~\text{mW}$ (blue dot-dashed line), $20~\text{mW}$ (green dotted line), $50~\text{mW}$ (purple dashed line) and $200~\text{mW}$ (red solid line). The force for each power has been normalised to unity for clarity, as it is only the axial equilibrium position, $z_{eq}$, that is of conern. (b) Variation in height above the water layer droplets of radius $2~\mu\text{m}$ (blue dotted line), $3~\mu\text{m}$ (red dot-dashed line), $4~\mu\text{m}$ (purple dashed line) and $5.5~\mu\text{m}$ (black solid line) are trapped as a function of trapping power (power gradients of Knox \textit{et al.}~\cite{Knox2007}). All but the $5.5~\mu\text{m}$ droplet curve stop due to the loss of axial equilibrium position at high powers as in (a). The objective axial displacement, $X=40~\mu\text{m}$, the water layer ($n_{w}=1.342$) is $10~\mu\text{m}$ thick, $\gamma=1$, $\theta_{0}=41.23^{\circ}$ and the coverslip refractive index $n_{g}=1.517$ for both (a) and (b).}
	\label{fig:pg}
	\end{center}
\end{figure}

We now examine if the theory predicts some of the behaviour observed during aerosol trapping experiments starting with points one and two in the introduction. In fig.~\ref{fig:pgA} we plot for a $4~\mu\text{m}$ water droplet, trapped in the experimental system depicted in fig.~\ref{fig:Wolf}, the predicted axial force curves for increasing trapping powers. Repeating for several droplet radii, the height above the water layer that a droplet is trapped, obtained from $z_{eq}$, can be plotted as a function of power as shown in fig.~\ref{fig:pgB}.

Figure~\ref{fig:pgA} successfully predicts two physical observations from experiments. As the trapping power increases the droplet equilibrium position, $z_{eq}$, and hence height above the underlying water layer, increases and with enough power eventually falls from the trap~\cite{BurnhamBrownian}. The curves of $2~\mu\text{m}$, $3~\mu\text{m}$ and $4~\mu\text{m}$ droplets in fig.~\ref{fig:pgB} do not continue for higher powers but the $5.5~\mu\text{m}$ droplet continues indefinitely. Clearly, if an equilibrium position exists in the efficiency curves alone, then the droplet will always remain trapped. If no such position exists then the force curve may eventually lose its equilibrium position with increasing power. This qualitatively explains our own experimentally observed results and the power gradients of Knox \textit{et al.}~\citep{Knox2007}, indicating their measured gradients~\citep{Knox2007} are a segment of the extended curves. With this enhanced understanding their suggestion of using power gradients as a tool for aerosol sizing could benefit.

Figure~\ref{fig:pgB} may also explain point four, why there is a linear dependence on captured droplet size as a function of power, and why small droplets cannot be trapped at high powers (figure 4 in~\cite{Burnham2006} and figure 2 in~\cite{McGloin2008}). The power gradients show that above certain powers, depending on droplet radius, it is possible for no equilibrium position to exist. Therefore, although a `large' droplet may be trapped at relatively large powers, smaller droplets cannot be for the same power.

\begin{figure}[!ht]
	\begin{center}
	\includegraphics[width=8.6cm]{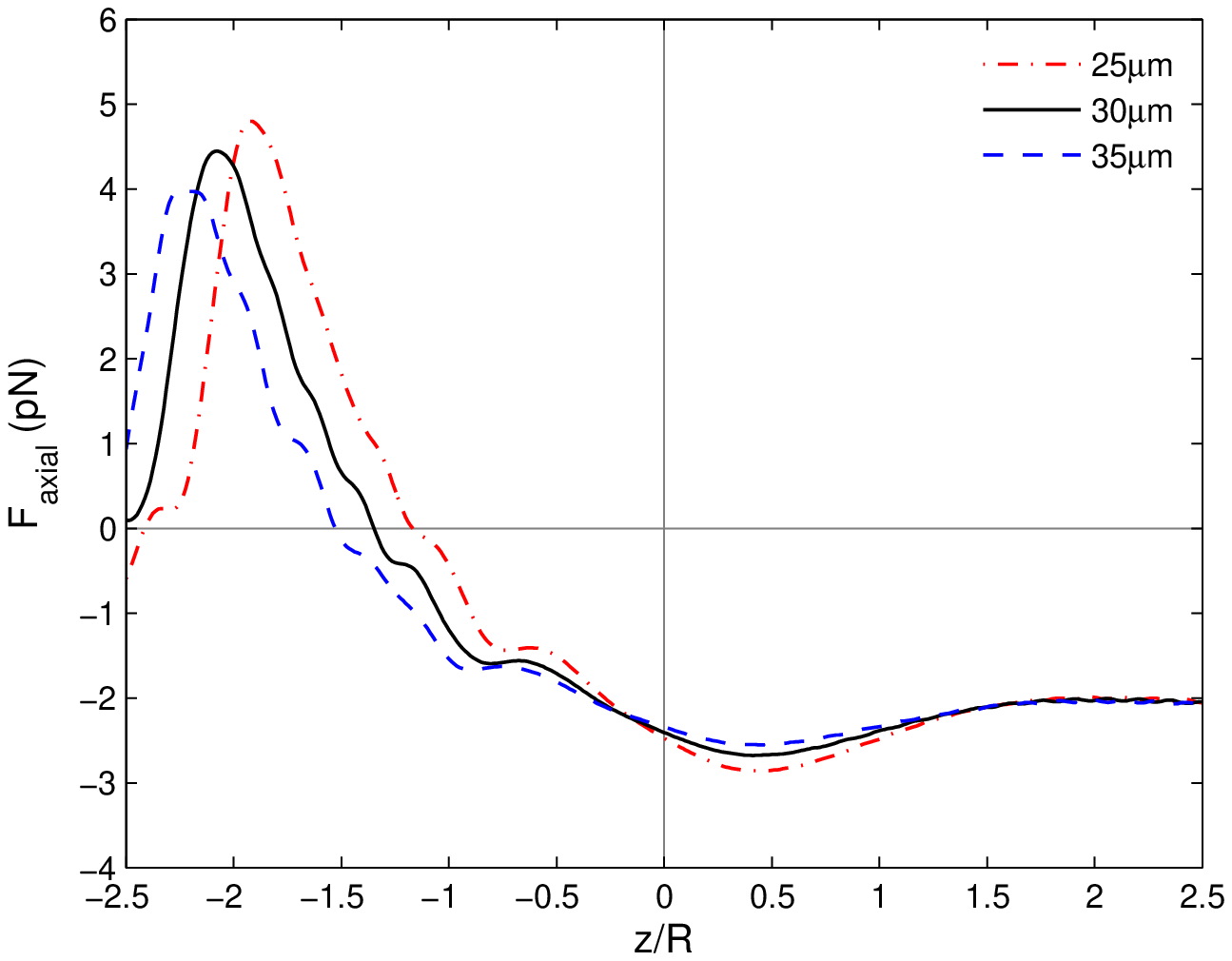}
	\caption{(Color online) Variation of axial force for a $4~\mu\text{m}$ water droplet ($n_{p}=1.342$) trapped in air ($n_{a}=1.000$) with $8~\text{mW}$ of power for microscope objective displacements of $25~\mu\text{m}$ (red dot-dashed line), $30~\mu\text{m}$ (solid black line) and $35~\mu\text{m}$ (dashed blue line). The water layer ($n_{w}=1.342$) is $10~\mu\text{m}$ thick, $\gamma=1$, $\theta_{0}=41.23^{\circ}$ and the coverslip refractive index $n_{g}=1.517$.}
	\label{fig:hg}
	\end{center}
\end{figure}

A large parameter that governs the magnitude of the spherical aberration induced by the interfaces is the depth into the sample which the beam is focussed. For example, a lower focus has less aberration. In fig.~\ref{fig:hg} the beam is simulated to focus at several depths into the sample chamber and the force curve calculated again for a $4~\mu\text{m}$ water droplet. 

The decrease in aberration not only shifts $z_{eq}$ closer to the paraxial focus but also increases the strength of the optical trap with an increasing $F_{z,max}^{-}$ and overall deepening of the potential well.

\begin{figure}[!ht]
	\begin{center}
	\includegraphics[width=8.6cm]{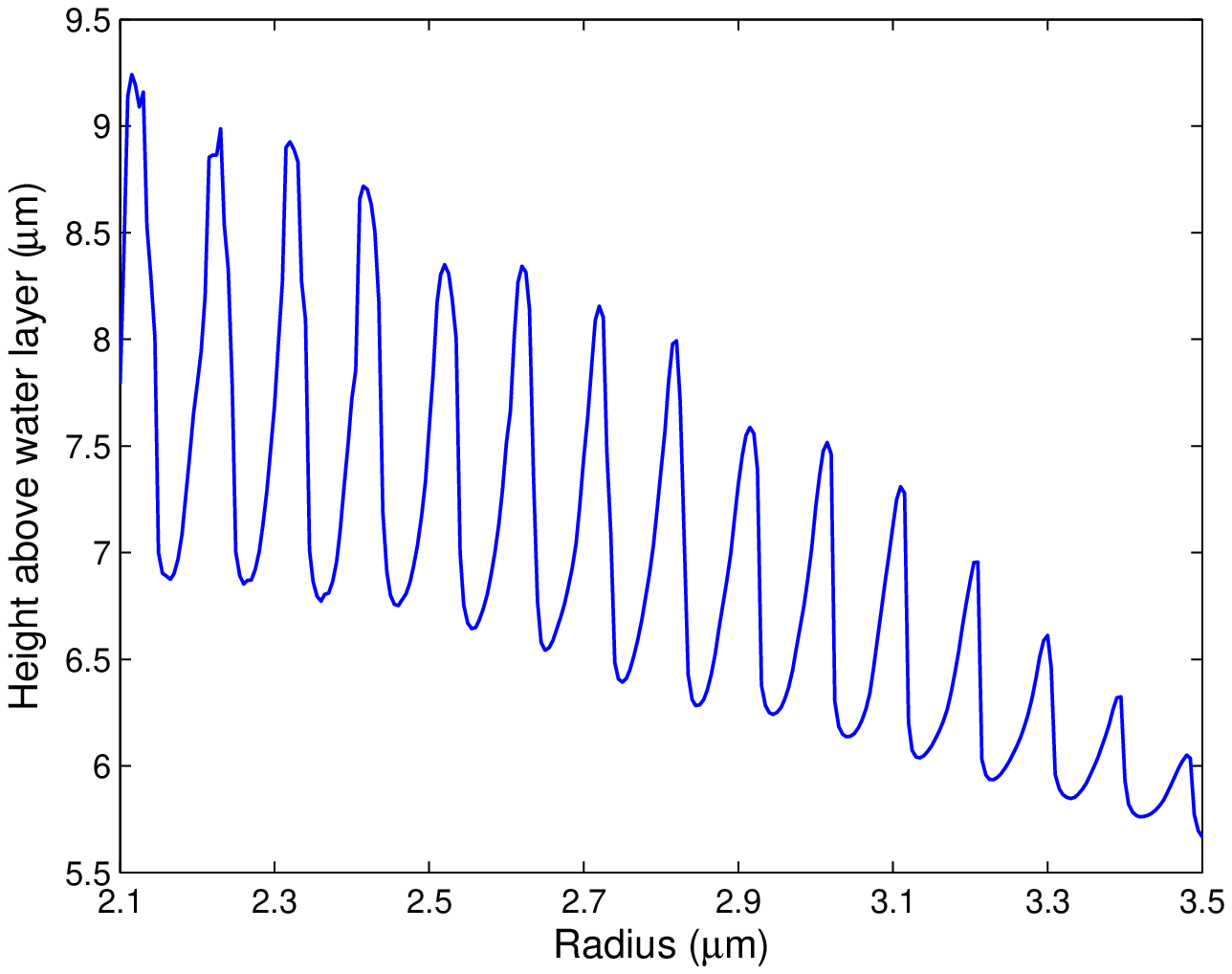}
	\caption{(Color online) Plot of the height a water droplet ($n_{p}=1.342$) in air ($n_{a}=1.000$) is trapped above the underlying water layer ($n_{w}=1.342$) as a function of radius. The objective axial displacement, $X=25~\mu\text{m}$, the water layer is $10~\mu\text{m}$ thick, $\gamma=1$, $\theta_{0}=41.23^{\circ}$, the coverslip refractive index $n_{g}=1.517$ and the trap power is $10~\text{mW}$}
	\label{fig:osc}
	\end{center}
\end{figure}

Now consider point three from the introduction. It is reasonable to expect the particle to exhibit some sort of interferometric properties with reflections from the inside of the droplet interfering with themselves. This can be demonstrated by simplifying the model and treating it as an interferometer which performs reasonably well at estimating the axial efficiency at the paraxial focus~\citep{Neto2000}. Liquid aerosols will establish a stable size once in equilibrium with their surrounding environment, namely the relative humidity~\cite{Mitchem2008}. Although the process of growth and evaporation is relatively fast it is at times clear one of these occurs just after the droplet becomes trapped. Investigating how the height at which the droplet is trapped varies with droplet radius we plot fig.~\ref{fig:osc}.

There is a clear, near sinusoidal, oscillation in droplet height as a function of its radius. A single oscillation in height occurs over a change in droplet radius of $\sim100~\text{nm}$, going from a local minima to maxima in half this, $\sim50~\text{nm}$. So, a change in trap height of $\sim2~\mu\text{m}$ occurs due to only a $50~\text{nm}$ change in droplet radius. When observing a particle just after capture the change in size is clear, far above the limit of resolution, so must be greater than $50~\text{nm}$. Knowing that the oscillations are most frequent just after capture we conclude the multiple oscillations that occur in experiments are due to changing particle radius and hence equilibrium position, $z_{eq}$.

To measure this oscillation would be a challenging experiment. The droplets would need to be imaged from the side to measure their height and also coupled with a high precision sizing technique such as CERS [ref]. With such instruments in place the droplet radius would have to be varied by altering ambient relative humidity or varying droplet temperature [ref]. Both would also change the salt concentration of the droplet hence refractive index so this effect would need to be included in the model.

\subsection{Limits of techniques}
\begin{figure}[!ht]
	\begin{center}
	\includegraphics[width=8.6cm]{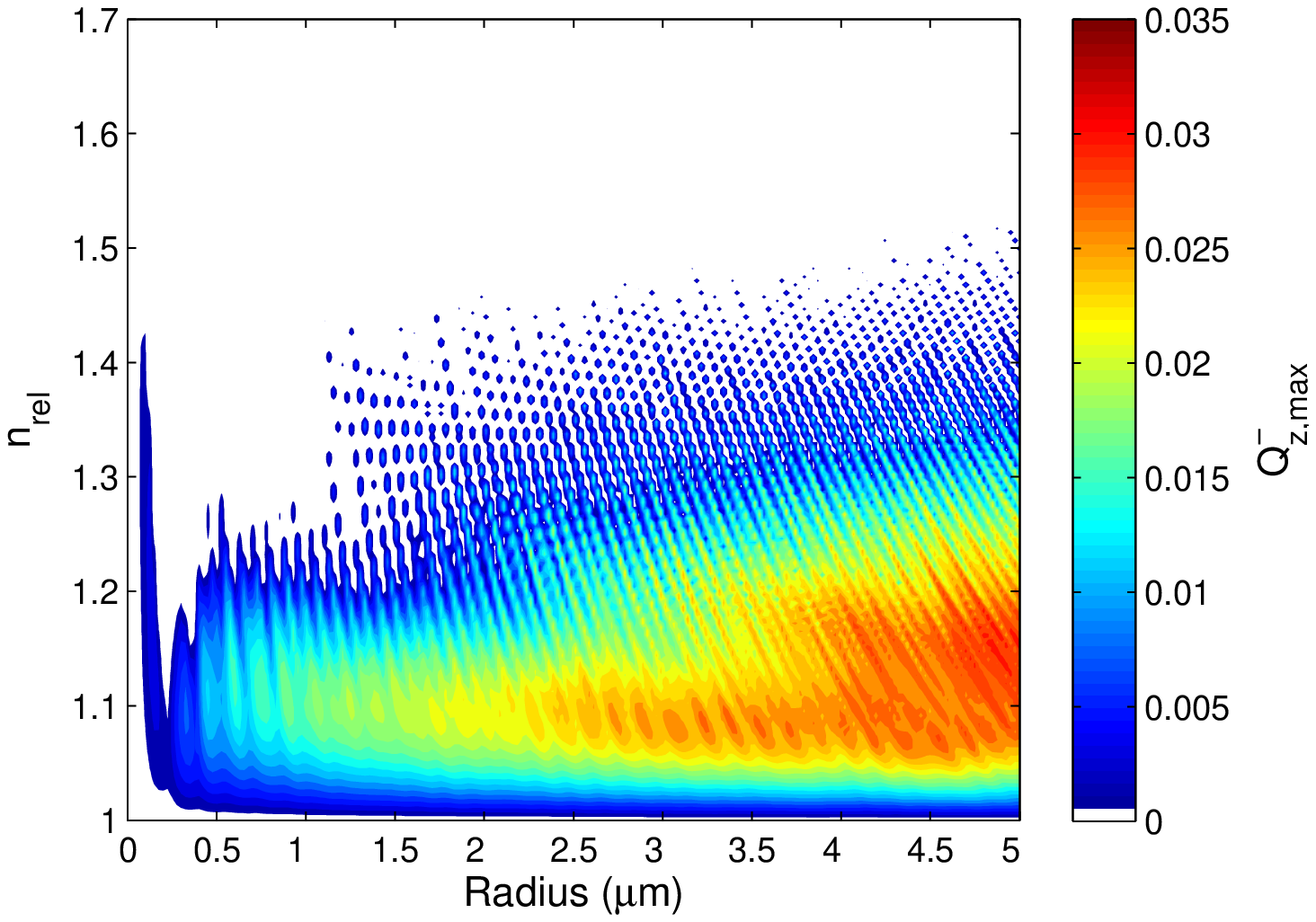}
	\caption{(Color online)$Q_{z,max}^{-}$ as a function of relative refractive index and radius for spheres trapped in a water medium ($n_{w}=1.33$). The objective axial displacement, $X=40~\mu\text{m}$, $\gamma=1$, $\theta_{0}=61.25^{\circ}$ and the coverslip refractive index $n_{g}=1.517$. The colour bar in (b) is representative for both plots. (a) is an expanded view of the first $2.5~\mu\text{m}$ of (b) as this is the region where the boundaries would really like to be pushed.}
	\label{fig:qwater}
	\end{center}
\end{figure}

\begin{figure}[!ht]
	\begin{center}
	\includegraphics[width=8.6cm]{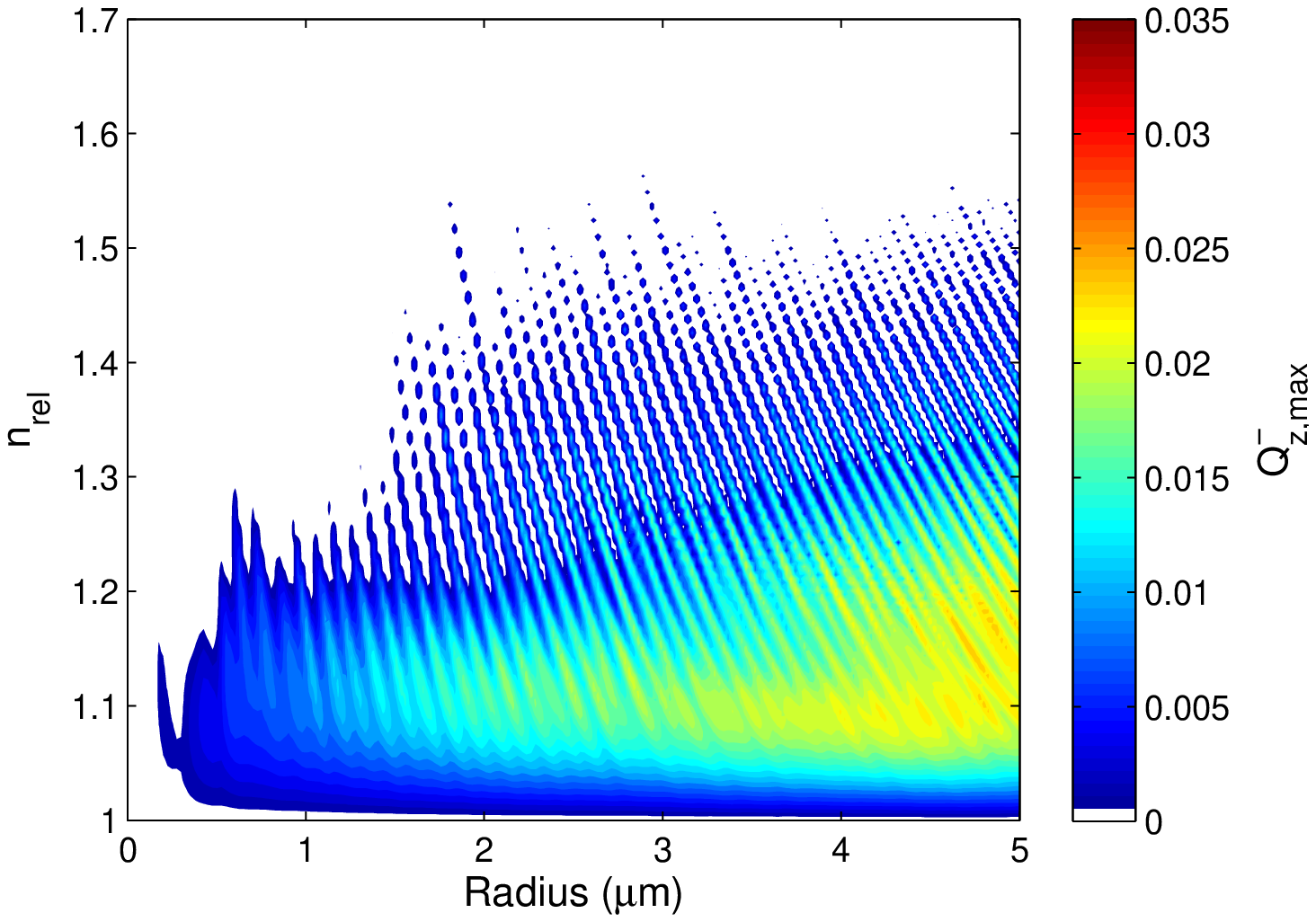}\label{fig:qairB}
	\caption{(Color online)$Q_{z,max}^{-}$ as a function of relative refractive index and radius for spheres trapped in an air medium ($n_{a}=1.000$). The objective axial displacement, $X=40~\mu\text{m}$, the water layer ($n_{w}=1.342$) is $10~\mu\text{m}$ thick, $\gamma=1$, $\theta_{0}=41.23^{\circ}$ and the coverslip refractive index $n_{g}=1.517$. The colour bar in (b) is representative for both plots. (a) Is an expanded view of the first $2.5~\mu\text{m}$ of (b) as this is the region where the boundaries would really like to be pushed.}
	\label{fig:qair}
	\end{center}
\end{figure}

Section B qualitatively explained the appearance of four unique phenomena observed when trapping airborne water droplets by simulating an experimental system through modelling the optical forces created by the focussing of a high NA beam through two refractive index mismatched interfaces. In what follows we shall see how far the boundaries of optical trapping in air can be pushed. We shall investigate if smaller particles ($<1~\mu\text{m}$) can be trapped, if the axial trap strength and capture volume can be increased, and determine the limits on the particle refractive index that can be trapped. To explore these questions $Q_{z,max}^{-}$ is calculated as a function of both particle radius and relative refractive index~\citep{Stilgoe2008}, first for particles suspended in water, fig.~\ref{fig:qwater}, as means of comparison, and then for airborne particles; fig.~\ref{fig:qair}.

The white areas on the plots represent parameter space where a negative $Q_{axial}$ value does not exist and hence no stable trap position is possible~\footnote[2]{This is not completely accurate as there is a very small value of $Q_{z,max}^{-}$ in the white areas but it is negligible~\citep{Sun2009,Nieminen2009}.}. Of immediate note are the `spikes' in the contour plots indicating resonances in the force experienced by the particles. The effect is more pronounced as a function of radius although at the high refractive index end of the spikes there are rapid resonances in force as a function of refractive index, creating tiny islands of parameter space where traps can exist.

These resonances can be explained by interference effects due to the spheres increased refelctivity at high relative refractive index and its associated variation with radius~\citep{Stilgoe2008}. The decreased frequency of the resonances in air, fig.~\ref{fig:qair}, is due to the lower medium refractive index ($n_{a}=1.00$ and $n_{w}=1.33$).

As the plots are functions of relative refractive index it is noted that for a given particle refractive index the horizontal line of interest is higher up the refractive index axis in air than water. Looking at the sorts of particles normally trapped in both media gives a `feel' for the plots. For example, a silica sphere in water exists along the line defined by $n_{rel}\simeq1.09$ in fig.~\ref{fig:qwater} and for a water droplet in air the line is at $n_{rel}\simeq1.34$ in fig.~\ref{fig:qair}.

The continuous region of stability for optical tweezers in air is over a smaller range of refractive indices ($\Delta n_{rel}\simeq1-1.25$) than when trapping in water ($\Delta n_{rel}\simeq1.33-1.65$) and, also, the maximum negative axial efficiency values, $Q_{z,max}^{-}$, are smaller overall for trapping in air than in water. This is understandable because the larger relative refractive index between particle and medium for aerosols increases the Fresnel reflection coefficients, hence increases scattering forces which probably overcome the gradient forces. The minimum radius possible to trap is smaller in water than air probably due to the increased spherical aberration in the focussed beam, induced by the coverslip interface, which has a larger refractive index contrast in airborne traps. As seen in fig.~\ref{fig:profiles} this increased aberration produces larger period oscillations in intensity allowing more `room' for particles to `fall' between.

\begin{figure}[!ht]
	\begin{center}
	\includegraphics[width=8.6cm]{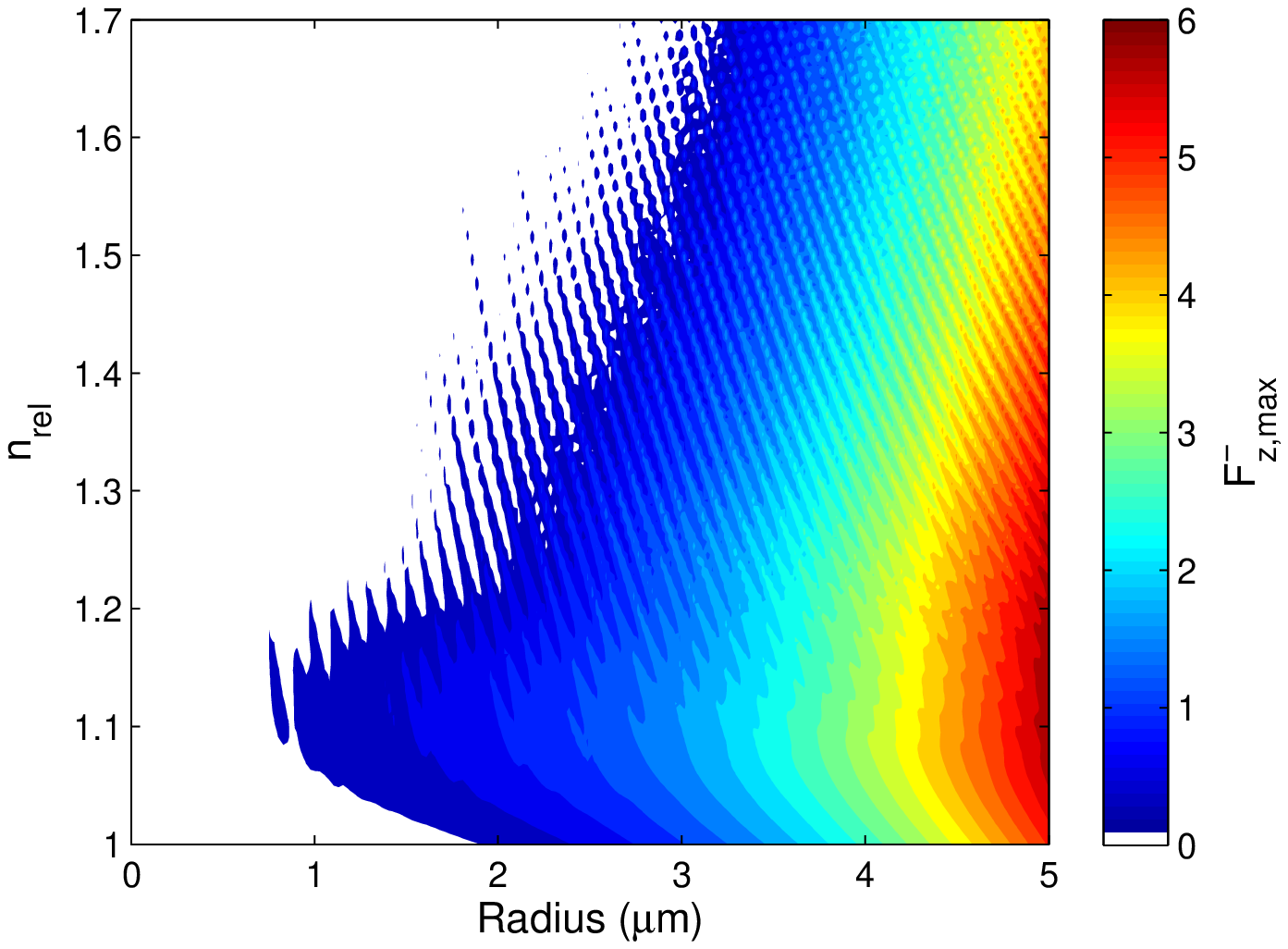}\label{fig:fairB}
	\caption{(Color online)$F_{z,max}^{-}$ as a function of relative refractive index and radius for spheres trapped in an air medium ($n_{a}=1.000$). The objective axial displacement, $X=40~\mu\text{m}$, the water layer ($n_{w}=1.342$) is $10~\mu\text{m}$ thick, $\gamma=1$, $\theta_{0}=41.23^{\circ}$ and the coverslip refractive index $n_{g}=1.517$. The colour bar in (b) is representative for both plots. (a) Is an expanded view of the first $2.5~\mu\text{m}$ of (b).}
	\label{fig:fair}
	\end{center}
\end{figure}

These plots are deceptive; the true range of particles that would theoretically obtain an axial equilibrium position in air has been misquoted because, as previously mentioned, the weight of the particle is significant. $F_{z,max}^{-}$ is the truly relevant quantity that will allow the determination of whether the spheres are isolated in three dimensions or not. Unfortunately, this poses a problem as the force from equation~\ref{eq:momentumefficiency} is dependent on laser power and with this additional variable not all parameter space can be easily displayed. Instead $F_{z,max}^{-}$ is plotted for a single power, $P=10~\text{mW}$, in fig.~\ref{fig:fair}.

Comparing figures~\ref{fig:qair} and~\ref{fig:fair} we come to an interesting conclusion. For droplets with certain particle parameters, indicated in fig.~\ref{fig:qair}, traps are created through the transfer of optical momentum alone (single beam gradient force trap or optical tweezers). However, fig.~\ref{fig:fair} indicates that with the assistance of gravity a larger range of droplets can be `trapped', although not with momentum transfer alone. Consider a droplet that evolves in size (it will also evolve slightly in refractive index due to salt concentration changes); as the radius varies the `path' of the particle in the parameter space of fig.~\ref{fig:qair} may cross through a non-tweezing region but due to its weight remains trapped (fig.~\ref{fig:fair}). This difficulty in deciding whether a droplet is tweezed or levitated leads to the conclusion that as a general name for the experimental tool being used we really have a \textit{quasi optical tweezers}.

\begin{figure}[!ht]
	\begin{center}
	\includegraphics[width=8cm]{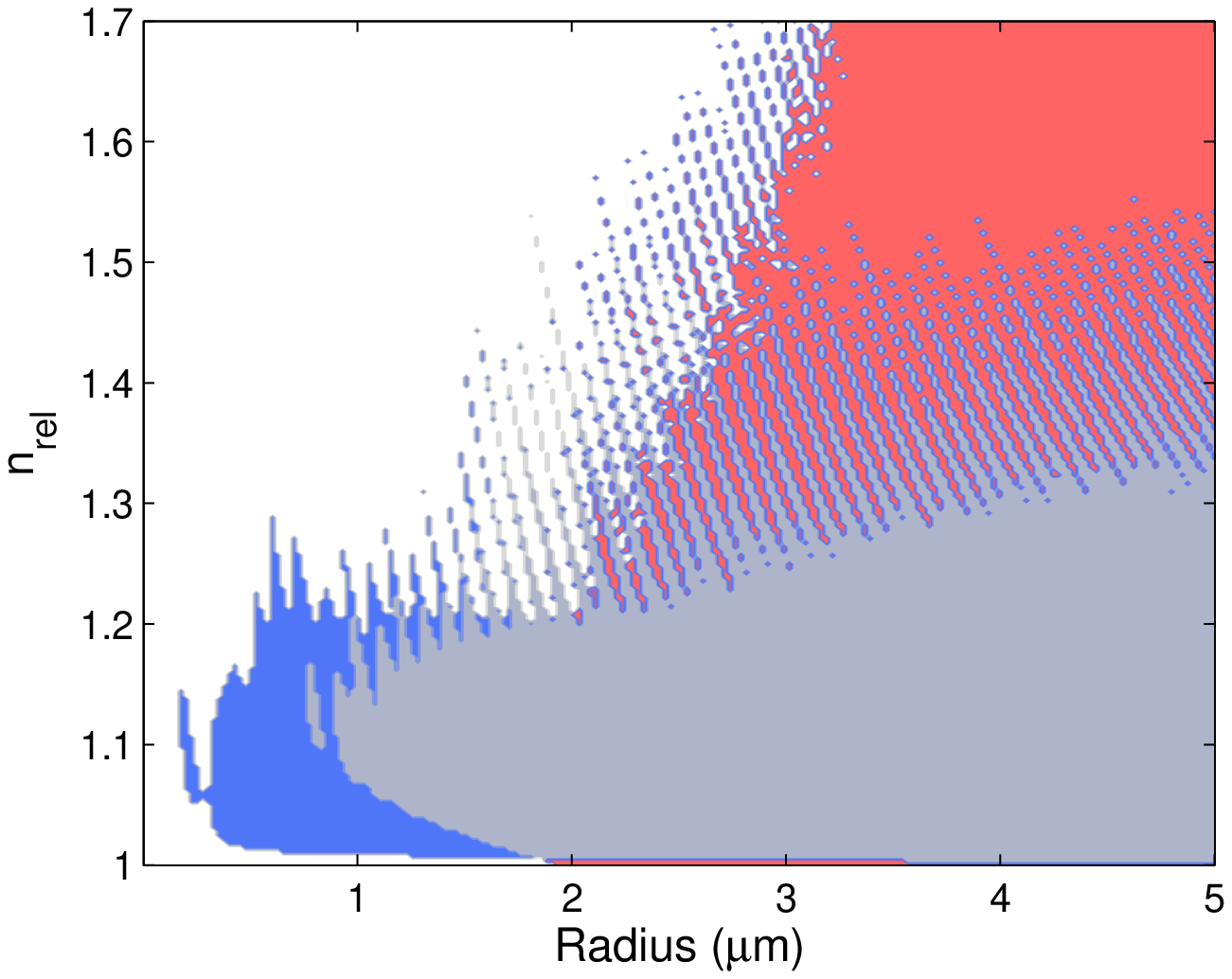}
	\caption{(Color online) Superposition of figures~\ref{fig:qair} and~\ref{fig:fair} highlighting the areas of parameter space, as a function of relative refractive index and radius, where water droplets are truly optically tweezed (high $R$, low $n_{rel}$, grey), only trapped with the assistance of gravity (high $R$, high $n_{rel}$, red), and optically tweezed if the droplet had neutral buoyancy (low $R$, low $n_{rel}$, blue). The white area represents areas where neither optical tweezing nor levitation occurs. The parameters for these plots are the same as the respective figures.}
	\label{fig:superimposed}
	\end{center}
\end{figure}

In fig.~\ref{fig:superimposed} we superimpose the tweezing and trapping areas of figures~\ref{fig:qair} and~\ref{fig:fair}. Areas of parameter space truly optically tweezed are coloured grey, areas that are only trapped with the assistance of gravity are coloured red and the area that would be truly tweezed if the droplets had neutral buoyancy is coloured blue. White areas retain the same meaning of neither tweezing nor levitation.

It is clear now that the choice of inverted or non-inverted tweezers is critical in the success of optically trapping a large range of aerosol sizes. Having established the true nature of the technique we are using the next section discusses the possibility of moving more of parameter space into the tweezing regime (trapping through momentum transfer alone).

\subsection{Optimisation and extension of limits}
We have demonstrated several points that stop airborne traps from reaching their optimum performance. These include spherical aberration created in the beam, a high refractive index contrast between particle and medium causing large scattering forces, and a lack of high converging angles (reduced NA) due to total internal reflection at the coverslip interface.

Total internal reflection is not easily circumvented but it could be possible to correct for spherical aberration or possibly remove the large scattering forces which we shall now explore.

\subsubsection{Spherical aberration correction}
\begin{figure}[!ht]
	\begin{center}
	\includegraphics[width=8.6cm]{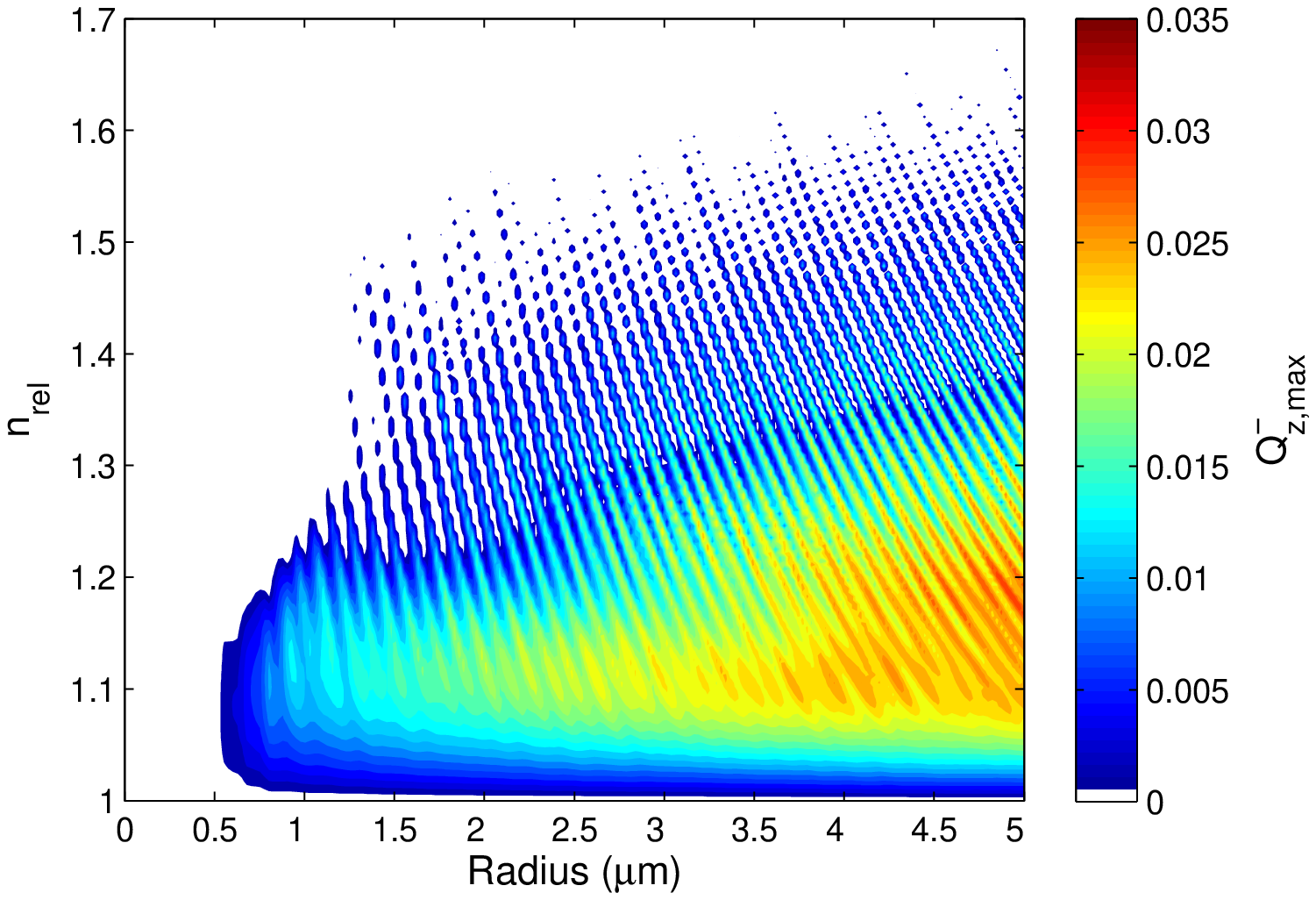}
	\caption{(Color online) $Q_{z,max}^{-}$ as a function of relative refractive index and radius for spheres trapped in an air medium ($n_{a}=1.00$) with a Gaussian beam entering the back aperture of the microscope objective with an additional spherical aberration placed on the beam at the entrance to the objective back aperture of magnitude $0.08\lambda$. The objective axial displacement, $X=40~\mu\text{m}$, the water layer is $10~\mu\text{m}$ ($n_{w}=1.342$) thick, $\gamma=1$, $\theta_{0}=41.23^{\circ}$ and the coverslip refractive index $n_{g}=1.517$. (a) is an expanded view of the first $2.5~\mu\text{m}$ of (b).}
	\label{fig:qairSA}
	\end{center}
\end{figure}

It is feasible to correct for the spherical aberration using spatial light modulators, deformable mirrors, or other wavefront modfiying elements~\cite{Theofanidou2004}. Any correction would clearly be advantageous creating better localisation of aerosols and hopefully moving into the important accumulation mode size regime. In fig.~\ref{fig:qairSA} we plot for the same parameters as fig.~\ref{fig:qair} but an additional spherical aberration, of magnitude $0.08\lambda$, is placed on the beam input to the objective.

Figure~\ref{fig:qairSA} shows an improvement in the axial strength of the optical tweezers, an increase in the overall range of parameter space that can be tweezed but unfortunately also an increase in the minimum particle radius that can be tweezed.

\subsubsection{Removal of beam centre intensity}
\begin{figure}[!ht]
	\begin{center}
	\includegraphics[width=8.6cm]{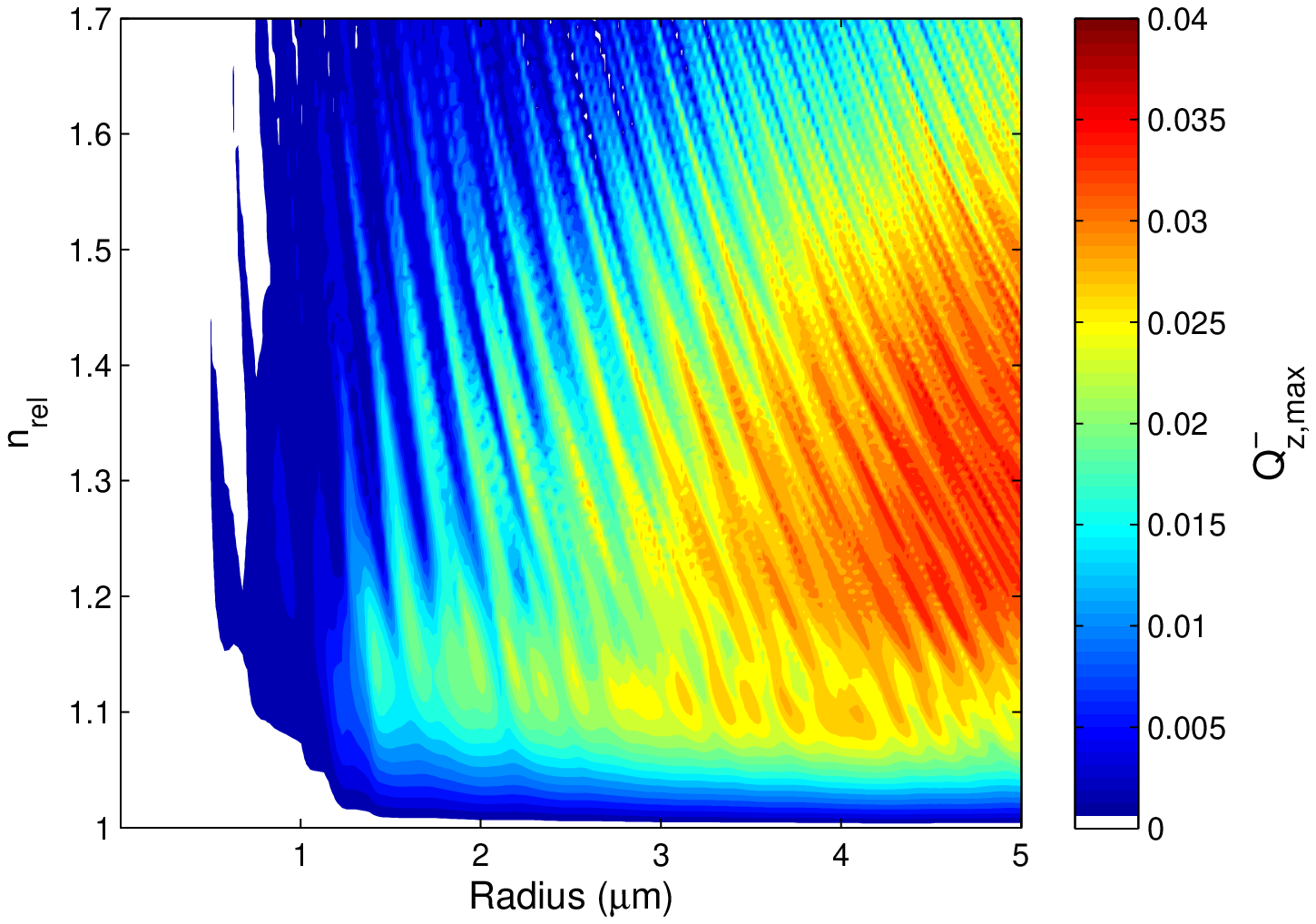}
	\caption{(Color online) $Q_{z,max}^{-}$ as a function of relative refractive index and radius for spheres trapped in an air medium ($n_{a}=1.000$) with a Gaussian beam entering the back aperture of the microscope objective with 57\% of its central area removed. The objective axial displacement, $X=40~\mu\text{m}$, the water layer ($n_{w}=1.342$) is $10~\mu\text{m}$ thick, $\gamma=1$, $\theta_{0}=41.23^{\circ}$ and the coverslip refractive index $n_{g}=1.517$.}
	\label{fig:qandfairwithoutcore}
	\end{center}
\end{figure}

Ashkin used GO to predict an increase in $Q_{z,max}^{-}$ if a `doughnut' mode beam ($\text{TEM}_{01}$) fills the back aperture of an objective. Increasing $Q_{z,max}^{-}$ is by far the most difficult problem in airborne tweezers, as shown and discussed, so next we predict the effects of removing the central portion of a Gaussian beam as it enters the objective. In fig.~\ref{fig:qandfairwithoutcore} we plot $Q_{z,max}^{-}$ against both radius and relative refractive index for a Gaussian beam where $\sim57\%$ of the beam area is removed leaving an annulus, yet keeping the total power the same.

Figure~\ref{fig:qandfairwithoutcore} shows that the area in parameter space over which a true optical tweezer can be created is greatly increased by removing the central core of a Gaussian beam, although the minimum sphere radius tweezable has increased. The minimum radius increase could be due to the zero intensity that may now exist in the focal plane of the tweezers into which a small enough sphere could sit, experiencing no forces from the surrounding light.

The predicted increase in parameter space over which aerosols can be tweezed is of great promise to the field of aerosol optical manipulation. It is difficult to trap high refractive index aerosols, specifically solid microspheres, yet they are of great importance to the fields of medicinal drug lung delivery and atmospheric chemistry ~\cite{Jacobson2002, Labiris2003}. It is hoped that a definitive experiment can be performed in future to verify this increase in optical tweezers parameter space.

\subsection{Capture volume}\label{sec:cv}
We have demonstrated that power gradients may explain the linear dependence of captured droplet size on trapping power but there may be more interactions occurring than thought. In order for the droplets to become trapped they must enter a capture volume, so it would be pertinent to calculate how this volume varies with trapping power and droplet radius. This volume extends between the maximum and minimum force points in the axial and lateral directions simultaneously. Unfortunately, evaluation of this volume requires calculation of forces for locations away from the optical axis. At these locations there is a complex interplay between axial and lateral efficiencies~\citep{Mazolli2003,Viana2007}. This will require more study to ascertain a suitable description and answer to the question of whether the capture volume plays a significant role in the linear dependence of captured droplet size on trapping power.

\section{Shortcomings of theory}
\begin{figure}[!ht]
	\begin{center}
	\includegraphics[width=8.6cm]{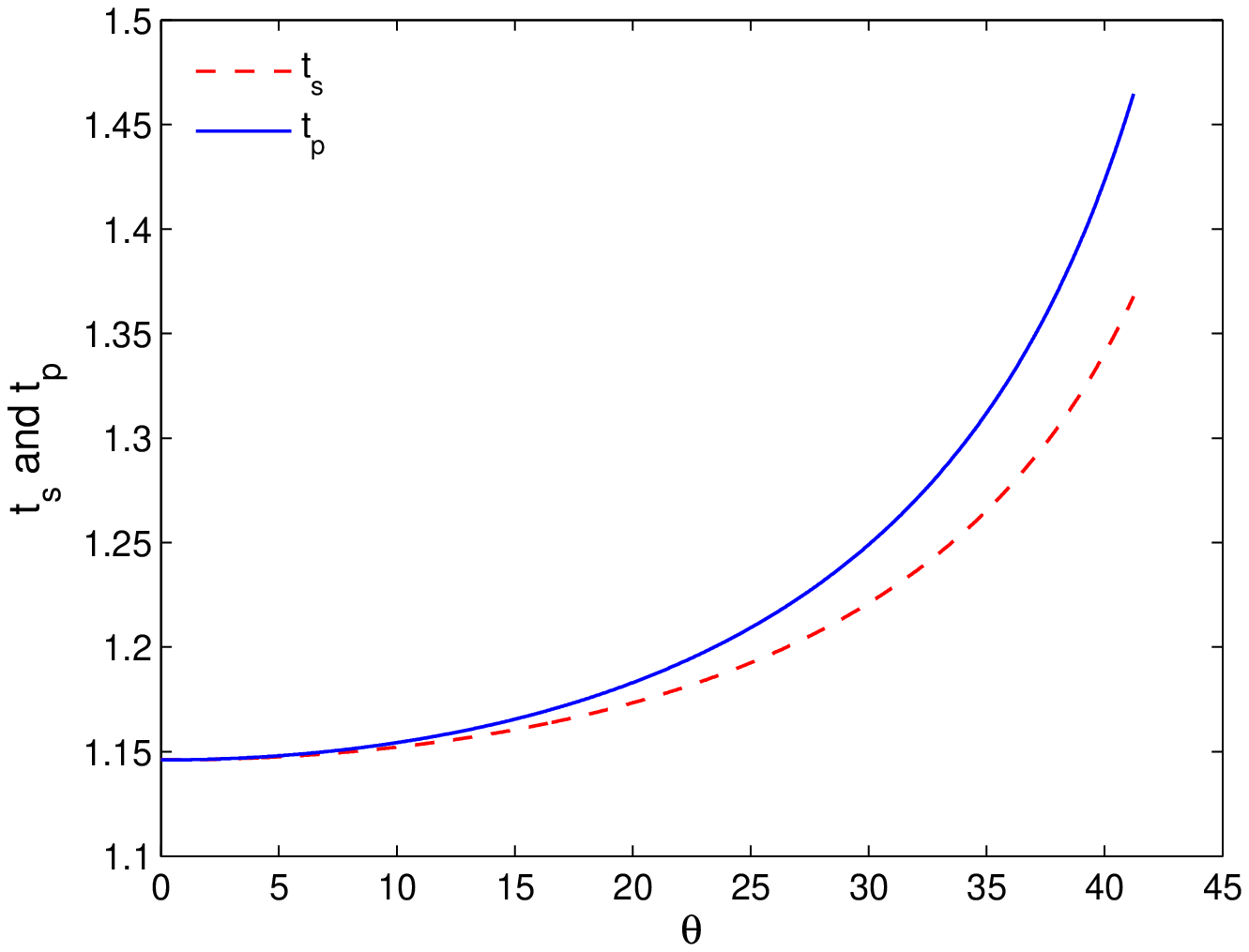}%
	\caption{(Color online) Fresnel transmission coefficients $t_{s}$ (black solid) and $t_{p}$ (red dashed) for TE and TM modes of polarisation respectively as a function of incident angle $\theta$ up to the critical angle $\simeq41.5^{\circ}$ for a glass ($n_{g}=1.517$) to air ($n_{a}=1.000$) refractive index interface.}
	\label{fig:T}
	\end{center}
\end{figure}

Within the Mie scattering theory outlined above we make the assumption that the Fresnel transmission coefficients, $t_{s}$ and $t_{p}$, for TE and TM modes of polarisation respectively are equal and take $t_{s}=t_{p}$. Plotted in fig.~\ref{fig:T} are both modes of the transmission coefficients as a function of incident angle on a glass to air interface to highlight the inaccuracy in this assumption.

The discrepancy in fig.~\ref{fig:T} is significant. We propose that in future work it should be attempted to place both TE and TM modes into the theory.

\section{Conclusion}
We have outlined a modified MDSA model of optical traps to describe our experimental system, and shown that it must be used over a simpler GO approximation in order to fully describe the beam profile and scattering. Investigation of the MDSA parameters indicate spherical aberration must be included, and we have discussed that gravity must not be na\"{i}vely ignored.

Once correctly described we use our extension to MDSA theory to qualitatively predict and explain four unusual phenomena observed only in airborne optical traps. It has also allowed us to explore limitations of current experiments and approaches before investigating what can be improved.

The resonance type plots go some way to explaining what the experimenter observes: it has been observed experimentally that water droplets ($n_{rel}\simeq1.342$) are easily trapped for a wide range of sizes, but we can say from experience that it is difficult to trap small water aerosols ($\leq1~\mu\text{m}$) even though they are produced from the nebuliser. In our previous work~\citep{Summers2008} where solid aerosols ($n_{rel}\simeq1.445$) are trapped, we can say that the expected range of particles trapped was odd. In colloidal systems, if two particle sizes can be trapped, almost certainly a size between these two will also be trapped. However, in air, spheres with certain radii could not be trapped yet sizes both above and below could.

Looking at fig.~\ref{fig:fair} both these phenomena can be qualitatively explained with the existence of the resonances as a function of radius and the lack of $Q_{z,max}^{-}$ for small spheres ($\leq1~\mu\text{m}$). Obviously, it is not easy to prove the non-result of being unable to trap certain objects, but the results here give some indication as to why it is so hard to trap small spheres, with a relatively high refractive index, that are so easily trapped in water.

This work has lead to many new insights into how aerosols are trapped in single beam gradient force traps. Namely that an optical trap in the inverted geometry behaves as a quasi optical tweezer, at times tweezing droplets and at others only levitating them. We have given qualitative predictions that explain physical phenomena observed experimentally and explored the theoretical limits of trapping aerosols, both which help to define the parameters of the current tools at our disposal. The challenge for the future is to produce quantitative agreement between experiment and theory.


\begin{acknowledgments}
DRB is a Lindemann Trust Fellow. DM is a Royal Society University Research Fellow.
\end{acknowledgments}

\bibliographystyle{apsrev}
\bibliography{DanBurnhamsBibliography16-01-10}

\begin{thebibliography}{55}
\expandafter\ifx\csname natexlab\endcsname\relax\def\natexlab#1{#1}\fi
\expandafter\ifx\csname bibnamefont\endcsname\relax
  \def\bibnamefont#1{#1}\fi
\expandafter\ifx\csname bibfnamefont\endcsname\relax
  \def\bibfnamefont#1{#1}\fi
\expandafter\ifx\csname citenamefont\endcsname\relax
  \def\citenamefont#1{#1}\fi
\expandafter\ifx\csname url\endcsname\relax
  \def\url#1{\texttt{#1}}\fi
\expandafter\ifx\csname urlprefix\endcsname\relax\def\urlprefix{URL }\fi
\providecommand{\bibinfo}[2]{#2}
\providecommand{\eprint}[2][]{\url{#2}}

\bibitem[{\citenamefont{Pesce et~al.}(2005)\citenamefont{Pesce, Sasso, and
  Fusco}}]{Pesce2005}
\bibinfo{author}{\bibfnamefont{G.}~\bibnamefont{Pesce}},
  \bibinfo{author}{\bibfnamefont{A.}~\bibnamefont{Sasso}}, \bibnamefont{and}
  \bibinfo{author}{\bibfnamefont{S.}~\bibnamefont{Fusco}},
  \bibinfo{journal}{Review of Scientific Instruments}
  \textbf{\bibinfo{volume}{76}}, \bibinfo{pages}{115105}
  (\bibinfo{year}{2005}).

\bibitem[{\citenamefont{Ghislain et~al.}(1994)\citenamefont{Ghislain, Switz,
  and Webb}}]{Ghislain1994}
\bibinfo{author}{\bibfnamefont{L.~P.} \bibnamefont{Ghislain}},
  \bibinfo{author}{\bibfnamefont{N.~A.} \bibnamefont{Switz}}, \bibnamefont{and}
  \bibinfo{author}{\bibfnamefont{W.~W.} \bibnamefont{Webb}},
  \bibinfo{journal}{Review of Scientific Instruments}
  \textbf{\bibinfo{volume}{65}}, \bibinfo{pages}{2762} (\bibinfo{year}{1994}).

\bibitem[{\citenamefont{Lang et~al.}(2002)\citenamefont{Lang, Asbury, Shaevitz,
  and Block}}]{Lang2002}
\bibinfo{author}{\bibfnamefont{M.~J.} \bibnamefont{Lang}},
  \bibinfo{author}{\bibfnamefont{C.~L.} \bibnamefont{Asbury}},
  \bibinfo{author}{\bibfnamefont{J.~W.} \bibnamefont{Shaevitz}},
  \bibnamefont{and} \bibinfo{author}{\bibfnamefont{S.~M.} \bibnamefont{Block}},
  \bibinfo{journal}{Biophysical Journal} \textbf{\bibinfo{volume}{83}},
  \bibinfo{pages}{491} (\bibinfo{year}{2002}).

\bibitem[{\citenamefont{Kuyper et~al.}(2003)\citenamefont{Kuyper, Brewood, and
  Chiu}}]{Kuyper2003}
\bibinfo{author}{\bibfnamefont{C.~L.} \bibnamefont{Kuyper}},
  \bibinfo{author}{\bibfnamefont{G.~P.} \bibnamefont{Brewood}},
  \bibnamefont{and} \bibinfo{author}{\bibfnamefont{D.~T.} \bibnamefont{Chiu}},
  \bibinfo{journal}{Nano Letters} \textbf{\bibinfo{volume}{3}},
  \bibinfo{pages}{1387} (\bibinfo{year}{2003}), ISSN \bibinfo{issn}{1530-6984}.

\bibitem[{\citenamefont{Di~Leonardo et~al.}(2006)\citenamefont{Di~Leonardo,
  Leach, Mushfique, Cooper, Ruocco, and Padgett}}]{Di2006}
\bibinfo{author}{\bibfnamefont{R.}~\bibnamefont{Di~Leonardo}},
  \bibinfo{author}{\bibfnamefont{J.}~\bibnamefont{Leach}},
  \bibinfo{author}{\bibfnamefont{H.}~\bibnamefont{Mushfique}},
  \bibinfo{author}{\bibfnamefont{J.~M.} \bibnamefont{Cooper}},
  \bibinfo{author}{\bibfnamefont{G.}~\bibnamefont{Ruocco}}, \bibnamefont{and}
  \bibinfo{author}{\bibfnamefont{M.~J.} \bibnamefont{Padgett}},
  \bibinfo{journal}{Physical Review Letters} \textbf{\bibinfo{volume}{96}},
  \bibinfo{pages}{134502} (\bibinfo{year}{2006}), ISSN
  \bibinfo{issn}{0031-9007}.

\bibitem[{\citenamefont{Hertlein et~al.}(2008)\citenamefont{Hertlein, Helden,
  Gambassi, Dietrich, and Bechinger}}]{Hertlein2008}
\bibinfo{author}{\bibfnamefont{C.}~\bibnamefont{Hertlein}},
  \bibinfo{author}{\bibfnamefont{L.}~\bibnamefont{Helden}},
  \bibinfo{author}{\bibfnamefont{A.}~\bibnamefont{Gambassi}},
  \bibinfo{author}{\bibfnamefont{S.}~\bibnamefont{Dietrich}}, \bibnamefont{and}
  \bibinfo{author}{\bibfnamefont{C.}~\bibnamefont{Bechinger}},
  \bibinfo{journal}{Nature} \textbf{\bibinfo{volume}{451}},
  \bibinfo{pages}{172} (\bibinfo{year}{2008}).

\bibitem[{\citenamefont{McCann et~al.}(1999)\citenamefont{McCann, Dykman, and
  Golding}}]{McCann1999}
\bibinfo{author}{\bibfnamefont{L.~I.} \bibnamefont{McCann}},
  \bibinfo{author}{\bibfnamefont{M.}~\bibnamefont{Dykman}}, \bibnamefont{and}
  \bibinfo{author}{\bibfnamefont{B.}~\bibnamefont{Golding}},
  \bibinfo{journal}{Nature} \textbf{\bibinfo{volume}{402}},
  \bibinfo{pages}{785} (\bibinfo{year}{1999}).

\bibitem[{\citenamefont{Mitchem and Reid}(2008)}]{Mitchem2008}
\bibinfo{author}{\bibfnamefont{L.}~\bibnamefont{Mitchem}} \bibnamefont{and}
  \bibinfo{author}{\bibfnamefont{J.~P.} \bibnamefont{Reid}},
  \bibinfo{journal}{Chemical Society Reviews} \textbf{\bibinfo{volume}{37}},
  \bibinfo{pages}{756} (\bibinfo{year}{2008}).

\bibitem[{\citenamefont{Zhao et~al.}(2008)\citenamefont{Zhao, Milne, Edgar,
  Jeffries, McGloin, and Chiu}}]{Zhao2008}
\bibinfo{author}{\bibfnamefont{Y.}~\bibnamefont{Zhao}},
  \bibinfo{author}{\bibfnamefont{G.}~\bibnamefont{Milne}},
  \bibinfo{author}{\bibfnamefont{J.~S.} \bibnamefont{Edgar}},
  \bibinfo{author}{\bibfnamefont{G.~D.~M.} \bibnamefont{Jeffries}},
  \bibinfo{author}{\bibfnamefont{D.}~\bibnamefont{McGloin}}, \bibnamefont{and}
  \bibinfo{author}{\bibfnamefont{D.~T.} \bibnamefont{Chiu}},
  \bibinfo{journal}{Applied Physics Letters} \textbf{\bibinfo{volume}{92}},
  \bibinfo{pages}{161111} (\bibinfo{year}{2008}).

\bibitem[{\citenamefont{Metzger et~al.}(2006)\citenamefont{Metzger, Wright, and
  Dholakia}}]{Metzger2006a}
\bibinfo{author}{\bibfnamefont{N.~K.} \bibnamefont{Metzger}},
  \bibinfo{author}{\bibfnamefont{E.~M.} \bibnamefont{Wright}},
  \bibnamefont{and} \bibinfo{author}{\bibfnamefont{K.}~\bibnamefont{Dholakia}},
  \bibinfo{journal}{New Journal of Physics} \textbf{\bibinfo{volume}{8}},
  \bibinfo{pages}{139} (\bibinfo{year}{2006}).

\bibitem[{\citenamefont{Knoner et~al.}(2006)\citenamefont{Knoner, Parkin,
  Nieminen, Heckenberg, and Rubinsztein-Dunlop}}]{Knoner2006}
\bibinfo{author}{\bibfnamefont{G.}~\bibnamefont{Knoner}},
  \bibinfo{author}{\bibfnamefont{S.}~\bibnamefont{Parkin}},
  \bibinfo{author}{\bibfnamefont{T.~A.} \bibnamefont{Nieminen}},
  \bibinfo{author}{\bibfnamefont{N.~R.} \bibnamefont{Heckenberg}},
  \bibnamefont{and}
  \bibinfo{author}{\bibfnamefont{H.}~\bibnamefont{Rubinsztein-Dunlop}},
  \bibinfo{journal}{Physical Review Letters} \textbf{\bibinfo{volume}{97}},
  \bibinfo{pages}{157402} (\bibinfo{year}{2006}).

\bibitem[{\citenamefont{Guck et~al.}(2005)\citenamefont{Guck, Schinkinger,
  Lincoln, Wottawah, Ebert, Romeyke, Lenz, Erickson, Ananthakrishnan, Mitchell
  et~al.}}]{Guck2005}
\bibinfo{author}{\bibfnamefont{J.}~\bibnamefont{Guck}},
  \bibinfo{author}{\bibfnamefont{S.}~\bibnamefont{Schinkinger}},
  \bibinfo{author}{\bibfnamefont{B.}~\bibnamefont{Lincoln}},
  \bibinfo{author}{\bibfnamefont{F.}~\bibnamefont{Wottawah}},
  \bibinfo{author}{\bibfnamefont{S.}~\bibnamefont{Ebert}},
  \bibinfo{author}{\bibfnamefont{M.}~\bibnamefont{Romeyke}},
  \bibinfo{author}{\bibfnamefont{D.}~\bibnamefont{Lenz}},
  \bibinfo{author}{\bibfnamefont{H.~M.} \bibnamefont{Erickson}},
  \bibinfo{author}{\bibfnamefont{R.}~\bibnamefont{Ananthakrishnan}},
  \bibinfo{author}{\bibfnamefont{D.}~\bibnamefont{Mitchell}},
  \bibnamefont{et~al.}, \bibinfo{journal}{Biophysical Journal}
  \textbf{\bibinfo{volume}{88}}, \bibinfo{pages}{3689} (\bibinfo{year}{2005}).

\bibitem[{\citenamefont{Reid}(2005)}]{Reid2005}
\bibinfo{author}{\bibnamefont{Reid}}, in \emph{\bibinfo{booktitle}{Colloid
  Science: Principles, Methods and Applications}}, edited by
  \bibinfo{editor}{\bibfnamefont{T.}~\bibnamefont{Cosgrove}}
  (\bibinfo{publisher}{Wiley-Blackwell}, \bibinfo{year}{2005}),
  vol.~\bibinfo{volume}{10}.

\bibitem[{\citenamefont{Jacobson}(2002)}]{Jacobson2002}
\bibinfo{author}{\bibfnamefont{M.~Z.} \bibnamefont{Jacobson}},
  \emph{\bibinfo{title}{Atmospheric Pollution: History, Science, and
  Regulation}} (\bibinfo{publisher}{Cambridge University Press},
  \bibinfo{year}{2002}).

\bibitem[{\citenamefont{Labiris and Dolovich}(2003)}]{Labiris2003}
\bibinfo{author}{\bibfnamefont{N.~R.} \bibnamefont{Labiris}} \bibnamefont{and}
  \bibinfo{author}{\bibfnamefont{M.~B.} \bibnamefont{Dolovich}},
  \bibinfo{journal}{British Journal of Clinical Pharmacology}
  \textbf{\bibinfo{volume}{56}}, \bibinfo{pages}{588} (\bibinfo{year}{2003}).

\bibitem[{\citenamefont{Smith et~al.}(2003)\citenamefont{Smith, Cui, and
  Bustamante}}]{Smith2003a}
\bibinfo{author}{\bibfnamefont{S.~B.} \bibnamefont{Smith}},
  \bibinfo{author}{\bibfnamefont{Y.~J.} \bibnamefont{Cui}}, \bibnamefont{and}
  \bibinfo{author}{\bibfnamefont{C.}~\bibnamefont{Bustamante}},
  \bibinfo{journal}{Methods in Enzymology} \textbf{\bibinfo{volume}{361}},
  \bibinfo{pages}{134} (\bibinfo{year}{2003}).

\bibitem[{\citenamefont{Nieminen et~al.}(2001)\citenamefont{Nieminen,
  Rubinsztein-Dunlop, Heckenberg, and Bishop}}]{Nieminen2001a}
\bibinfo{author}{\bibfnamefont{T.~A.} \bibnamefont{Nieminen}},
  \bibinfo{author}{\bibfnamefont{H.}~\bibnamefont{Rubinsztein-Dunlop}},
  \bibinfo{author}{\bibfnamefont{N.~R.} \bibnamefont{Heckenberg}},
  \bibnamefont{and} \bibinfo{author}{\bibfnamefont{A.~I.}
  \bibnamefont{Bishop}}, \bibinfo{journal}{Computer Physics Communications}
  \textbf{\bibinfo{volume}{142}}, \bibinfo{pages}{468} (\bibinfo{year}{2001}).

\bibitem[{\citenamefont{Nieminen et~al.}(2007)\citenamefont{Nieminen, Loke,
  Stilgoe, Knoner, Branczyk, Heckenberg, and
  Rubinsztein-Dunlop}}]{Nieminen2007}
\bibinfo{author}{\bibfnamefont{T.~A.} \bibnamefont{Nieminen}},
  \bibinfo{author}{\bibfnamefont{V.~L.~Y.} \bibnamefont{Loke}},
  \bibinfo{author}{\bibfnamefont{A.~B.} \bibnamefont{Stilgoe}},
  \bibinfo{author}{\bibfnamefont{G.}~\bibnamefont{Knoner}},
  \bibinfo{author}{\bibfnamefont{A.~M.} \bibnamefont{Branczyk}},
  \bibinfo{author}{\bibfnamefont{N.~R.} \bibnamefont{Heckenberg}},
  \bibnamefont{and}
  \bibinfo{author}{\bibfnamefont{H.}~\bibnamefont{Rubinsztein-Dunlop}},
  \bibinfo{journal}{Journal of Optics A - Pure and Applied Optics}
  \textbf{\bibinfo{volume}{9}}, \bibinfo{pages}{S196} (\bibinfo{year}{2007}).

\bibitem[{\citenamefont{Viana et~al.}(2007)\citenamefont{Viana, Rocha,
  Mesquita, Mazolli, Neto, and Nussenzveig}}]{Viana2007}
\bibinfo{author}{\bibfnamefont{N.~B.} \bibnamefont{Viana}},
  \bibinfo{author}{\bibfnamefont{M.~S.} \bibnamefont{Rocha}},
  \bibinfo{author}{\bibfnamefont{O.~N.} \bibnamefont{Mesquita}},
  \bibinfo{author}{\bibfnamefont{A.}~\bibnamefont{Mazolli}},
  \bibinfo{author}{\bibfnamefont{P.~A.~M.} \bibnamefont{Neto}},
  \bibnamefont{and} \bibinfo{author}{\bibfnamefont{H.~M.}
  \bibnamefont{Nussenzveig}}, \bibinfo{journal}{Physical Review E}
  \textbf{\bibinfo{volume}{75}}, \bibinfo{pages}{021914}
  (\bibinfo{year}{2007}).

\bibitem[{\citenamefont{Ashkin et~al.}(1986)\citenamefont{Ashkin, Dziedzic,
  Bjorkholm, and Chu}}]{Ashkin1986}
\bibinfo{author}{\bibfnamefont{A.}~\bibnamefont{Ashkin}},
  \bibinfo{author}{\bibfnamefont{J.~M.} \bibnamefont{Dziedzic}},
  \bibinfo{author}{\bibfnamefont{J.~E.} \bibnamefont{Bjorkholm}},
  \bibnamefont{and} \bibinfo{author}{\bibfnamefont{S.}~\bibnamefont{Chu}},
  \bibinfo{journal}{Optics Letters} \textbf{\bibinfo{volume}{11}},
  \bibinfo{pages}{288} (\bibinfo{year}{1986}).

\bibitem[{\citenamefont{Roosen}(1977)}]{Roosen1977}
\bibinfo{author}{\bibfnamefont{G.}~\bibnamefont{Roosen}},
  \bibinfo{journal}{Optics Communications} \textbf{\bibinfo{volume}{21}},
  \bibinfo{pages}{189} (\bibinfo{year}{1977}).

\bibitem[{\citenamefont{Burnham and McGloin}()}]{BurnhamBrownian}
\bibinfo{author}{\bibfnamefont{D.~R.} \bibnamefont{Burnham}} \bibnamefont{and}
  \bibinfo{author}{\bibfnamefont{D.}~\bibnamefont{McGloin}},
  \bibinfo{journal}{submitted}  (????).

\bibitem[{\citenamefont{Knox et~al.}(2007)\citenamefont{Knox, Reid, Hanford,
  Hudson, and Mitchem}}]{Knox2007}
\bibinfo{author}{\bibfnamefont{K.~J.} \bibnamefont{Knox}},
  \bibinfo{author}{\bibfnamefont{J.~P.} \bibnamefont{Reid}},
  \bibinfo{author}{\bibfnamefont{K.~L.} \bibnamefont{Hanford}},
  \bibinfo{author}{\bibfnamefont{A.~J.} \bibnamefont{Hudson}},
  \bibnamefont{and} \bibinfo{author}{\bibfnamefont{L.}~\bibnamefont{Mitchem}},
  \bibinfo{journal}{Journal of Optics A - Pure and Applied Optics}
  \textbf{\bibinfo{volume}{9}}, \bibinfo{pages}{S180} (\bibinfo{year}{2007}).

\bibitem[{\citenamefont{McGloin et~al.}(2008)\citenamefont{McGloin, Burnham,
  Summers, Rudd, Dewar, and Anand}}]{McGloin2008}
\bibinfo{author}{\bibfnamefont{D.}~\bibnamefont{McGloin}},
  \bibinfo{author}{\bibfnamefont{D.~R.} \bibnamefont{Burnham}},
  \bibinfo{author}{\bibfnamefont{M.~D.} \bibnamefont{Summers}},
  \bibinfo{author}{\bibfnamefont{D.}~\bibnamefont{Rudd}},
  \bibinfo{author}{\bibfnamefont{N.}~\bibnamefont{Dewar}}, \bibnamefont{and}
  \bibinfo{author}{\bibfnamefont{S.}~\bibnamefont{Anand}},
  \bibinfo{journal}{Faraday Discussions} \textbf{\bibinfo{volume}{137}},
  \bibinfo{pages}{335} (\bibinfo{year}{2008}).

\bibitem[{\citenamefont{Di~Leonardo et~al.}(2007)\citenamefont{Di~Leonardo,
  Ruocco, Leach, Padgett, Wright, Girkin, Burnham, and McGloin}}]{Di2007a}
\bibinfo{author}{\bibfnamefont{R.}~\bibnamefont{Di~Leonardo}},
  \bibinfo{author}{\bibfnamefont{G.}~\bibnamefont{Ruocco}},
  \bibinfo{author}{\bibfnamefont{J.}~\bibnamefont{Leach}},
  \bibinfo{author}{\bibfnamefont{M.~J.} \bibnamefont{Padgett}},
  \bibinfo{author}{\bibfnamefont{A.~J.} \bibnamefont{Wright}},
  \bibinfo{author}{\bibfnamefont{J.~M.} \bibnamefont{Girkin}},
  \bibinfo{author}{\bibfnamefont{D.~R.} \bibnamefont{Burnham}},
  \bibnamefont{and} \bibinfo{author}{\bibfnamefont{D.}~\bibnamefont{McGloin}},
  \bibinfo{journal}{Physical Review Letters} \textbf{\bibinfo{volume}{99}},
  \bibinfo{pages}{010601} (\bibinfo{year}{2007}).

\bibitem[{\citenamefont{Hopkins et~al.}(2004)\citenamefont{Hopkins, Mitchem,
  Ward, and Reid}}]{Hopkins2004}
\bibinfo{author}{\bibfnamefont{R.~J.} \bibnamefont{Hopkins}},
  \bibinfo{author}{\bibfnamefont{L.}~\bibnamefont{Mitchem}},
  \bibinfo{author}{\bibfnamefont{A.~D.} \bibnamefont{Ward}}, \bibnamefont{and}
  \bibinfo{author}{\bibfnamefont{J.~P.} \bibnamefont{Reid}},
  \bibinfo{journal}{Physical Chemistry Chemical Physics}
  \textbf{\bibinfo{volume}{6}}, \bibinfo{pages}{4924} (\bibinfo{year}{2004}).

\bibitem[{\citenamefont{Burnham and McGloin}(2006)}]{Burnham2006}
\bibinfo{author}{\bibfnamefont{D.~R.} \bibnamefont{Burnham}} \bibnamefont{and}
  \bibinfo{author}{\bibfnamefont{D.}~\bibnamefont{McGloin}},
  \bibinfo{journal}{Optics Express} \textbf{\bibinfo{volume}{14}},
  \bibinfo{pages}{4175} (\bibinfo{year}{2006}).

\bibitem[{\citenamefont{Stilgoe et~al.}(2008)\citenamefont{Stilgoe, Nieminen,
  Knoner, Heckenberg, and Rubinsztein-Dunlop}}]{Stilgoe2008}
\bibinfo{author}{\bibfnamefont{A.~B.} \bibnamefont{Stilgoe}},
  \bibinfo{author}{\bibfnamefont{T.~A.} \bibnamefont{Nieminen}},
  \bibinfo{author}{\bibfnamefont{G.}~\bibnamefont{Knoner}},
  \bibinfo{author}{\bibfnamefont{N.~R.} \bibnamefont{Heckenberg}},
  \bibnamefont{and}
  \bibinfo{author}{\bibfnamefont{H.}~\bibnamefont{Rubinsztein-Dunlop}},
  \bibinfo{journal}{Optics Express} \textbf{\bibinfo{volume}{16}},
  \bibinfo{pages}{15039} (\bibinfo{year}{2008}).

\bibitem[{\citenamefont{Joykutty et~al.}(2005)\citenamefont{Joykutty, Mathur,
  Venkataraman, and Natarajan}}]{Joykutty2005}
\bibinfo{author}{\bibfnamefont{J.}~\bibnamefont{Joykutty}},
  \bibinfo{author}{\bibfnamefont{V.}~\bibnamefont{Mathur}},
  \bibinfo{author}{\bibfnamefont{V.}~\bibnamefont{Venkataraman}},
  \bibnamefont{and}
  \bibinfo{author}{\bibfnamefont{V.}~\bibnamefont{Natarajan}},
  \bibinfo{journal}{Physical Review Letters} \textbf{\bibinfo{volume}{95}},
  \bibinfo{pages}{193902} (\bibinfo{year}{2005}).

\bibitem[{\citenamefont{Keen et~al.}(2007)\citenamefont{Keen, Leach, Gibson,
  and Padgett}}]{Keen2007}
\bibinfo{author}{\bibfnamefont{S.}~\bibnamefont{Keen}},
  \bibinfo{author}{\bibfnamefont{J.}~\bibnamefont{Leach}},
  \bibinfo{author}{\bibfnamefont{G.}~\bibnamefont{Gibson}}, \bibnamefont{and}
  \bibinfo{author}{\bibfnamefont{M.~J.} \bibnamefont{Padgett}},
  \bibinfo{journal}{Journal of Optics A - Pure and Applied Optics}
  \textbf{\bibinfo{volume}{9}}, \bibinfo{pages}{S264} (\bibinfo{year}{2007}).

\bibitem[{\citenamefont{Richards and Wolf}(1959)}]{Richards1959}
\bibinfo{author}{\bibfnamefont{B.}~\bibnamefont{Richards}} \bibnamefont{and}
  \bibinfo{author}{\bibfnamefont{E.}~\bibnamefont{Wolf}},
  \bibinfo{journal}{Proceedings of the Royal Society of London. Series A,
  Mathematical and Physical Sciences} \textbf{\bibinfo{volume}{253}},
  \bibinfo{pages}{358} (\bibinfo{year}{1959}).

\bibitem[{\citenamefont{Wolf}(1959)}]{Wolf1959}
\bibinfo{author}{\bibfnamefont{E.}~\bibnamefont{Wolf}},
  \bibinfo{journal}{Proceedings of the Royal Society of London. Series A,
  Mathematical and Physical Sciences} \textbf{\bibinfo{volume}{253}},
  \bibinfo{pages}{349} (\bibinfo{year}{1959}).

\bibitem[{\citenamefont{Kerker}(1969)}]{Kerker1969}
\bibinfo{author}{\bibfnamefont{M.}~\bibnamefont{Kerker}},
  \emph{\bibinfo{title}{The scattering of light, and other electromagnetic
  radiation}} (\bibinfo{year}{1969}).

\bibitem[{\citenamefont{Gong et~al.}(2007)\citenamefont{Gong, Wang, Li, Lou,
  and Xu}}]{Gong2007}
\bibinfo{author}{\bibfnamefont{Z.}~\bibnamefont{Gong}},
  \bibinfo{author}{\bibfnamefont{Z.}~\bibnamefont{Wang}},
  \bibinfo{author}{\bibfnamefont{Y.~M.} \bibnamefont{Li}},
  \bibinfo{author}{\bibfnamefont{L.~R.} \bibnamefont{Lou}}, \bibnamefont{and}
  \bibinfo{author}{\bibfnamefont{S.~H.} \bibnamefont{Xu}},
  \bibinfo{journal}{Optics Communications} \textbf{\bibinfo{volume}{273}},
  \bibinfo{pages}{37} (\bibinfo{year}{2007}).

\bibitem[{\citenamefont{Gussgard et~al.}(1992)\citenamefont{Gussgard, Lindmo,
  and Brevik}}]{Gussgard1992}
\bibinfo{author}{\bibfnamefont{R.}~\bibnamefont{Gussgard}},
  \bibinfo{author}{\bibfnamefont{T.}~\bibnamefont{Lindmo}}, \bibnamefont{and}
  \bibinfo{author}{\bibfnamefont{I.}~\bibnamefont{Brevik}},
  \bibinfo{journal}{Journal of The Optical Society of America B - Optical
  Physics} \textbf{\bibinfo{volume}{9}}, \bibinfo{pages}{1922}
  (\bibinfo{year}{1992}), ISSN \bibinfo{issn}{0740-3224}.

\bibitem[{\citenamefont{Kohira et~al.}(2005)\citenamefont{Kohira, Isomura,
  Magome, Mukai, and Yoshikawa}}]{Kohira2005}
\bibinfo{author}{\bibfnamefont{M.~I.} \bibnamefont{Kohira}},
  \bibinfo{author}{\bibfnamefont{A.}~\bibnamefont{Isomura}},
  \bibinfo{author}{\bibfnamefont{N.}~\bibnamefont{Magome}},
  \bibinfo{author}{\bibfnamefont{S.}~\bibnamefont{Mukai}}, \bibnamefont{and}
  \bibinfo{author}{\bibfnamefont{K.}~\bibnamefont{Yoshikawa}},
  \bibinfo{journal}{Chemical Physics Letters} \textbf{\bibinfo{volume}{414}},
  \bibinfo{pages}{389} (\bibinfo{year}{2005}), ISSN \bibinfo{issn}{0009-2614}.

\bibitem[{\citenamefont{Mazolli et~al.}(2003)\citenamefont{Mazolli, Neto, and
  Nussenzveig}}]{Mazolli2003}
\bibinfo{author}{\bibfnamefont{A.}~\bibnamefont{Mazolli}},
  \bibinfo{author}{\bibfnamefont{P.~A.~M.} \bibnamefont{Neto}},
  \bibnamefont{and} \bibinfo{author}{\bibfnamefont{H.~M.}
  \bibnamefont{Nussenzveig}}, \bibinfo{journal}{Proceedings of the Royal
  Society of London. Series A - Mathematical, Physical and Engineering
  Sciences} \textbf{\bibinfo{volume}{459}}, \bibinfo{pages}{3021}
  (\bibinfo{year}{2003}).

\bibitem[{\citenamefont{Torok and Varga}(1997)}]{Torok1997}
\bibinfo{author}{\bibfnamefont{P.}~\bibnamefont{Torok}} \bibnamefont{and}
  \bibinfo{author}{\bibfnamefont{P.}~\bibnamefont{Varga}},
  \bibinfo{journal}{Applied Optics} \textbf{\bibinfo{volume}{36}},
  \bibinfo{pages}{2305} (\bibinfo{year}{1997}).

\bibitem[{\citenamefont{Torok et~al.}(1995)\citenamefont{Torok, Varga, Laczik,
  and Booker}}]{Torok1995}
\bibinfo{author}{\bibfnamefont{P.}~\bibnamefont{Torok}},
  \bibinfo{author}{\bibfnamefont{P.}~\bibnamefont{Varga}},
  \bibinfo{author}{\bibfnamefont{Z.}~\bibnamefont{Laczik}}, \bibnamefont{and}
  \bibinfo{author}{\bibfnamefont{G.~R.} \bibnamefont{Booker}},
  \bibinfo{journal}{Journal of the Optical Society of America A - Optics Image
  Science and Vision} \textbf{\bibinfo{volume}{12}}, \bibinfo{pages}{325}
  (\bibinfo{year}{1995}).

\bibitem[{\citenamefont{Milne et~al.}(2007)\citenamefont{Milne, Dholakia,
  McGloin, Volke-Sepulveda, and Zemanek}}]{Milne2007a}
\bibinfo{author}{\bibfnamefont{G.}~\bibnamefont{Milne}},
  \bibinfo{author}{\bibfnamefont{K.}~\bibnamefont{Dholakia}},
  \bibinfo{author}{\bibfnamefont{D.}~\bibnamefont{McGloin}},
  \bibinfo{author}{\bibfnamefont{K.}~\bibnamefont{Volke-Sepulveda}},
  \bibnamefont{and} \bibinfo{author}{\bibfnamefont{P.}~\bibnamefont{Zemanek}},
  \bibinfo{journal}{Optics Express} \textbf{\bibinfo{volume}{15}},
  \bibinfo{pages}{13972} (\bibinfo{year}{2007}).

\bibitem[{\citenamefont{Bohren and Huffman}(1983)}]{Bohren1983}
\bibinfo{author}{\bibfnamefont{C.~F.} \bibnamefont{Bohren}} \bibnamefont{and}
  \bibinfo{author}{\bibfnamefont{D.~R.} \bibnamefont{Huffman}},
  \emph{\bibinfo{title}{Absorption and scattering of light by small particles}}
  (\bibinfo{publisher}{Wiley}, \bibinfo{year}{1983}).

\bibitem[{\citenamefont{Barton et~al.}(1989)\citenamefont{Barton, Alexander,
  and Schaub}}]{Barton1989}
\bibinfo{author}{\bibfnamefont{J.~P.} \bibnamefont{Barton}},
  \bibinfo{author}{\bibfnamefont{D.~R.} \bibnamefont{Alexander}},
  \bibnamefont{and} \bibinfo{author}{\bibfnamefont{S.~A.}
  \bibnamefont{Schaub}}, \bibinfo{journal}{Journal of Applied Physics}
  \textbf{\bibinfo{volume}{66}}, \bibinfo{pages}{4594} (\bibinfo{year}{1989}).

\bibitem[{\citenamefont{Born and Wolf}(1980)}]{Born1980}
\bibinfo{author}{\bibfnamefont{M.}~\bibnamefont{Born}} \bibnamefont{and}
  \bibinfo{author}{\bibfnamefont{E.}~\bibnamefont{Wolf}},
  \emph{\bibinfo{title}{Principles of Optics}} (\bibinfo{year}{1980}).

\bibitem[{\citenamefont{Farsund and Felderhof}(1996)}]{Farsund1996}
\bibinfo{author}{\bibfnamefont{O.}~\bibnamefont{Farsund}} \bibnamefont{and}
  \bibinfo{author}{\bibfnamefont{B.~U.} \bibnamefont{Felderhof}},
  \bibinfo{journal}{Physica A} \textbf{\bibinfo{volume}{227}},
  \bibinfo{pages}{108} (\bibinfo{year}{1996}).

\bibitem[{\citenamefont{Debye}(1909)}]{Debye1909}
\bibinfo{author}{\bibfnamefont{P.}~\bibnamefont{Debye}},
  \bibinfo{journal}{Annalen der Physik} \textbf{\bibinfo{volume}{335}},
  \bibinfo{pages}{57} (\bibinfo{year}{1909}).

\bibitem[{\citenamefont{Blanco et~al.}(1997)\citenamefont{Blanco, Florez, and
  Bermejo}}]{Blanco1997}
\bibinfo{author}{\bibfnamefont{M.~A.} \bibnamefont{Blanco}},
  \bibinfo{author}{\bibfnamefont{M.}~\bibnamefont{Florez}}, \bibnamefont{and}
  \bibinfo{author}{\bibfnamefont{M.}~\bibnamefont{Bermejo}},
  \bibinfo{journal}{Journal of Molecular Structure: THEOCHEM}
  \textbf{\bibinfo{volume}{419}}, \bibinfo{pages}{19} (\bibinfo{year}{1997}).

\bibitem[{\citenamefont{Edmonds}(1957)}]{Edmonds1957}
\bibinfo{author}{\bibfnamefont{A.~R.} \bibnamefont{Edmonds}},
  \emph{\bibinfo{title}{Angular momentum in quantum mechanics}}
  (\bibinfo{publisher}{Princeton University Press}, \bibinfo{year}{1957}).

\bibitem[{\citenamefont{Dutra et~al.}(2007)\citenamefont{Dutra, Viana, Neto,
  and Nussenzveig}}]{Dutra2007}
\bibinfo{author}{\bibfnamefont{R.~S.} \bibnamefont{Dutra}},
  \bibinfo{author}{\bibfnamefont{N.~B.} \bibnamefont{Viana}},
  \bibinfo{author}{\bibfnamefont{P.~A.~M.} \bibnamefont{Neto}},
  \bibnamefont{and} \bibinfo{author}{\bibfnamefont{H.~M.}
  \bibnamefont{Nussenzveig}}, \bibinfo{journal}{Journal of Optics A - Pure and
  Applied Optics} \textbf{\bibinfo{volume}{9}}, \bibinfo{pages}{S221}
  (\bibinfo{year}{2007}).

\bibitem[{\citenamefont{Neto and Nussenzveig}(2000)}]{Neto2000}
\bibinfo{author}{\bibfnamefont{P.~A.~M.} \bibnamefont{Neto}} \bibnamefont{and}
  \bibinfo{author}{\bibfnamefont{H.~M.} \bibnamefont{Nussenzveig}},
  \bibinfo{journal}{Europhysics Letters} \textbf{\bibinfo{volume}{50}},
  \bibinfo{pages}{702} (\bibinfo{year}{2000}).

\bibitem[{\citenamefont{F\"{a}llman and Axner}(2003)}]{Fallman2003}
\bibinfo{author}{\bibfnamefont{E.}~\bibnamefont{F\"{a}llman}} \bibnamefont{and}
  \bibinfo{author}{\bibfnamefont{O.}~\bibnamefont{Axner}},
  \bibinfo{journal}{Applied Optics} \textbf{\bibinfo{volume}{42}},
  \bibinfo{pages}{3915} (\bibinfo{year}{2003}).

\bibitem[{\citenamefont{Im et~al.}(2003)\citenamefont{Im, Kim, Joo, Oh, Song,
  Kim, and Park}}]{Im2003}
\bibinfo{author}{\bibfnamefont{K.~B.} \bibnamefont{Im}},
  \bibinfo{author}{\bibfnamefont{H.~I.} \bibnamefont{Kim}},
  \bibinfo{author}{\bibfnamefont{I.~J.} \bibnamefont{Joo}},
  \bibinfo{author}{\bibfnamefont{C.~H.} \bibnamefont{Oh}},
  \bibinfo{author}{\bibfnamefont{S.~H.} \bibnamefont{Song}},
  \bibinfo{author}{\bibfnamefont{P.~S.} \bibnamefont{Kim}}, \bibnamefont{and}
  \bibinfo{author}{\bibfnamefont{B.~C.} \bibnamefont{Park}},
  \bibinfo{journal}{Optics Communications} \textbf{\bibinfo{volume}{226}},
  \bibinfo{pages}{25} (\bibinfo{year}{2003}).

\bibitem[{\citenamefont{Theofanidou et~al.}(2004)\citenamefont{Theofanidou,
  Wilson, Hossack, and Arlt}}]{Theofanidou2004}
\bibinfo{author}{\bibfnamefont{E.}~\bibnamefont{Theofanidou}},
  \bibinfo{author}{\bibfnamefont{L.}~\bibnamefont{Wilson}},
  \bibinfo{author}{\bibfnamefont{W.~J.} \bibnamefont{Hossack}},
  \bibnamefont{and} \bibinfo{author}{\bibfnamefont{J.}~\bibnamefont{Arlt}},
  \bibinfo{journal}{Optics Communications} \textbf{\bibinfo{volume}{236}},
  \bibinfo{pages}{145} (\bibinfo{year}{2004}).

\bibitem[{\citenamefont{Summers et~al.}(2008)\citenamefont{Summers, Burnham,
  and McGloin}}]{Summers2008}
\bibinfo{author}{\bibfnamefont{M.~D.} \bibnamefont{Summers}},
  \bibinfo{author}{\bibfnamefont{D.~R.} \bibnamefont{Burnham}},
  \bibnamefont{and} \bibinfo{author}{\bibfnamefont{D.}~\bibnamefont{McGloin}},
  \bibinfo{journal}{Optics Express} \textbf{\bibinfo{volume}{16}},
  \bibinfo{pages}{7739} (\bibinfo{year}{2008}).

\bibitem[{\citenamefont{Sun and Grier}(2009)}]{Sun2009}
\bibinfo{author}{\bibfnamefont{B.}~\bibnamefont{Sun}} \bibnamefont{and}
  \bibinfo{author}{\bibfnamefont{D.~G.} \bibnamefont{Grier}},
  \bibinfo{journal}{Optics Express} \textbf{\bibinfo{volume}{17}},
  \bibinfo{pages}{2658} (\bibinfo{year}{2009}).

\bibitem[{\citenamefont{Nieminen et~al.}(2009)\citenamefont{Nieminen, Stilgoe,
  Loke, Heckenberg, and Rubinsztein-Dunlop}}]{Nieminen2009}
\bibinfo{author}{\bibfnamefont{T.~A.} \bibnamefont{Nieminen}},
  \bibinfo{author}{\bibfnamefont{A.~B.} \bibnamefont{Stilgoe}},
  \bibinfo{author}{\bibfnamefont{V.~L.~Y.} \bibnamefont{Loke}},
  \bibinfo{author}{\bibfnamefont{N.~R.} \bibnamefont{Heckenberg}},
  \bibnamefont{and}
  \bibinfo{author}{\bibfnamefont{H.}~\bibnamefont{Rubinsztein-Dunlop}},
  \bibinfo{journal}{Optics Express} \textbf{\bibinfo{volume}{17}},
  \bibinfo{pages}{2661} (\bibinfo{year}{2009}).

\end{thebibliography}


\end{document}